\documentclass{aa}
\usepackage[pdftex]{graphicx}
\usepackage{epsfig}
\usepackage{amsmath}
\usepackage{wasysym}

\usepackage{marvosym}
\usepackage{txfonts}
\pdfoutput=1
\usepackage[pdftex]{color,xcolor}
\usepackage[pdftex]{graphicx}
\usepackage{hyperref}
\hypersetup{pdfauthor=Christoph Mordasini}
\hypersetup{backref=true, pagebackref=true, hyperindex=true, breaklinks=true,colorlinks=true,urlcolor=blue, linkcolor=blue,  citecolor=blue,pagecolor=red, bookmarks=true, bookmarksopen=true}
\usepackage{epstopdf}

\usepackage{natbib}
\bibpunct{(}{)}{;}{a}{}{,} % to follow the A&A style
\usepackage{mathtools}

\def\mearth{M_\oplus}

\def\rhill{R_{\rm H}}
\def\mcore{M_{\rm core}}

\def\f1{f_{\rm I}}

\def\miso{M_{\rm iso}}

\def\beq{\begin{equation}}
\def\eeq{\end{equation}}

\def\fopa{f_{\rm opa}}
\def\t2{\tau_{\rm II}}

\def\sigmas0{\Sigma_{\rm s,0}}
\def\mj{M_{\textrm{\tiny \jupiter }}}

\newcommand{\rj}{R_{\textrm{\tiny \jupiter}}}
\def\tform{t_{\rm form}}
\def\s0{S_{\rm 0}}

\def\lacc{L_{\rm acc}}
\def\lint{L_{\rm int}}
\def\tKH{\tau_{\rm KH}}
\def\tacc{\tau_{\rm acc}}

\newcommand{\lsun}{L_{\odot}}

\newcommand{\mdotxy}{\dot{M}_{\rm XY}}

\defcitealias{marleyfortney2007}{M07}
\defcitealias{spiegelburrows2012}{SB12}

\def\({\left(}
\def\){\right)}
\def\<{\left<}
\def\>{\right>}

\begin{document}

\title{Luminosity of young Jupiters revisited} 
\subtitle{Massive cores make hot planets}
\author{C. Mordasini\thanks{Reimar-L\"ust Fellow of the MPG}}
\institute{Max-Planck-Institut f\"ur Astronomie, K\"onigstuhl 17, D-69117 Heidelberg, Germany} 
\offprints{Christoph MORDASINI, \email{mordasini@mpia.de}}
\date{Received 1.4.2013 / Accepted 22.6.2013}

\abstract
{The intrinsic luminosity of young Jupiters is of high interest for planet formation theory. It is an observable quantity that is determined by important physical mechanisms during formation, namely, the structure of the accretion shock, and even more fundamentally, the basic formation mechanism (core accretion or gravitational instability). }{ {Our aim is} to study the impact of the core mass on the post-formation entropy and luminosity of young giant planets forming via core accretion with a supercritical accretion shock that radiates all accretion shock energy (cold accretion).}{ {For this,} we conduct self-consistently coupled formation and evolution  calculations of  giant planets with masses between 1 and 12 Jovian masses and core masses between 20 and 120 Earth masses in the 1D spherically symmetric approximation.} { {As the main result, it is found} that the post-formation luminosity of massive giant planets is very sensitive to the core mass. An increase of the core mass by a factor 6 results in an increase of the post-formation luminosity of a 10 Jovian mass planet by a factor 120, indicating a dependency as $\mcore^{2-3}$.  Due to this dependency, there is no single well defined post-formation luminosity for core accretion, but a wide range, even for completely cold accretion. For massive cores  ($\sim$100 Earth masses), the post-formation luminosities of core accretion planets become so high that they approach those in the hot start scenario that is often associated with gravitational instability. For the mechanism to work, it is necessary that the solids are accreted before or during gas runaway accretion, and that they sink during this time deep into the planet.}{We make no claims whether or not such massive cores can actually form in giant planets especially at large orbital distances. But if they can form, it becomes difficult to rule out core accretion as formation mechanism based solely on luminosity  for directly imaged planets that are more luminous than predicted for low core masses. Instead of invoking gravitational instability as the consequently necessary formation mode, the high luminosity can also be caused, at least in principle, simply by a more massive core.}

\titlerunning{Luminosity of young Jupiters revisited}
\authorrunning{C. Mordasini}

\maketitle

\section{Introduction}
Young self-luminous giant planets are important for planet formation theory because their luminosity still depends on important physical mechanisms occurring during the precedent formation phase. It has previously been understood \citep[][hereafter \citetalias{marleyfortney2007}]{marleyfortney2007} that the structure of the accretion shock through which gas is accreted into a planet has very important consequences for the post-formation luminosity. Inspired by the stellar case \citep{stahlershu1980}, \citetalias[][]{marleyfortney2007} showed for planets forming via core accretion that a supercritical shock radiating away all liberated gravitational potential energy leads to low post-formation entropies and luminosities \citep[so-called cold accretion leading to a  cold start, e.g.,][]{hartmanncassen1997,baraffechabrier2009}. 

These luminosities are so much lower than in conventional so-called hot start simulations beginning with an arbitrary high entropy state \citep[e.g.,][]{burrowsmarley1997,baraffechabrier2003} that the difference is  easily observable with direct imaging observations. High entropy states are expected if the accretion shock is subcritical and does not radiate the shock luminosity (hot accretion), or if the  gas is incorporated into the planet without going through a shock at all.  This could happen if a large patch of the protoplanetary disk  becomes  collectively self-gravitationally unstable and directly forms a bound clump as envisioned in the  gravitational instability formation model. The results of \citet{galvagnihayfield2012} indicate that young giant planets formed by this mechanism indeed have high entropies.

One must, however, be careful in directly equating core accretion with cold start, and gravitational instability with hot start, as discussed in \citet{mordasinialibert2012}. In this work it was found that planets forming via the core accretion mechanism, but without radiative losses at the shock (hot accretion) have the same high post-formation entropies as typical for hot starts. Either completely cold, or completely hot accretion are limiting assumptions that are solely chosen due to the current lack of knowledge about the actual shock structure in the planetary case \citep[\citetalias{marleyfortney2007}, see][for the stellar first-core accretion shock]{commerconaudit2011}. It is certainly  reasonable to assume that the actual accretion into a planet (both for core accretion and gravitational instability) is neither completely hot nor completely cold. This motivated \citet[][hereafter \citetalias{spiegelburrows2012}]{spiegelburrows2012} to study a whole range of warm starts with intermediate entropies. Nevertheless, the potential difference in the luminosity of giant planets forming via the two different formation mechanisms explains why the luminosity of young planets is of highest interest for formation theory.

The ultimate goal must be to calculate the post-formation entropy for both mechanism with 3D radiation-hydrodynamic simulation that resolve the accretion geometry onto the planet (presumably via a circumplanetary disk), the shock physics and the transport of radiation. In this work we consider in contrast the strongly idealized case of 1D spherically symmetric accretion. But we are nevertheless able to point out a new effect that potentially makes that even for completely cold accretion of the gas, there could be young giant planets formed via core accretion that are almost as hot as in the conventional hot start scenario. 

The structure of the paper is as follows: in Section \ref{sect:model} we give an overview of the model and the settings used in the combined formation and evolution simulations that are presented in Sect. \ref{sect:simulations}. The resulting post-formation entropies are shown in Sect. \ref{sect:entropy} where we present an updated version of the ``entropy tuning fork''. In Section \ref{sect:luminosities} we show the luminosities during the evolution at constant mass  after formation while Section \ref{sect:radii} gives the radii during this phase. The conclusions are presented in Section \ref{sect:conclusions}.  {In the Appendices \ref{sect:l0mcore} and \ref{sect:modelsettingsMcoreS0} we give  further information about the power-law approximation for the post-formation luminosity and the impact of several model settings on the core mass - initial entropy mapping, respectively.}
 
\section{Model}\label{sect:model}
We simulate the formation and evolution of giant planets for different values of the initial surface density of planetesimals $\sigmas0$ that translate into different final core masses. We use the planet formation code based on the core accretion paradigm initially presented in \citet*{alibertmordasini2004} that was recently extended into a self-consistently coupled formation and evolution code as shown in \citet{mordasinialibert2012a} where a detailed description can be found. The model solves a slightly simplified version of the standard equations describing the 1D (radial) internal structure of giant planets, consisting of the equations of mass conservation, hydrostatic equilibrium, energy conservation and energy transport \citep[e.g.,][]{bodenheimerpollack1986}. The internal structure calculation is coupled with a model for the accretion rate of planetesimals as in \citet{pollackhubickyj1996} and the impact of planetesimals into the gaseous envelope \citep{podolakpollack1988,mordasinialibert2006}. In summary, the model allows us to make simulations  similar to those of \citet{fortneymarley2005a} or \citetalias{marleyfortney2007} that yield in particular the post-formation luminosity (i.e., the one after the final mass has been reached) of giant planets based self-consistently on the precedent formation phase.  

\subsection{Simulation setup}
We simulate the in situ formation and evolution of giant planets at 5.2 AU from a solar-like star.  Table \ref{tab:evocalcs} gives an overview of the most important settings and assumptions. Disk evolution and orbital migration are not included, but this should not alter the basic conclusions of the paper. The burning of deuterium is also neglected since it is not important at the maximal mass of 12 $\mj$   {(Jovian masses)} that we consider \citep{spiegelburrows2011,mollieremordasini2012,bodenheimerdangelo2013}. The pressure and temperature in the protoplanetary disk are the same as in the J1 simulation of \citet{pollackhubickyj1996}.  The opacity of grains is reduced by a factor 0.003 relative to the ISM to approximate the consequences of grain evolution \citep{movshovitzbodenheimer2010}. We do not model the hydrodynamic processes that govern the disk-limited gas accretion rate in the runaway phase $\dot{M}_{\rm XY,max}$  but fix it to  0.01 $\mearth$/yr, a typical value \citep{lissauerhubickyj2009}. We consider final total masses between 1 and 12 $\mj$, and vary the initial local surface density of planetesimals $\sigmas0$ so that planets with different core masses $\mcore$ form. The effect of planetesimal ejection is included, and the density of the core is variable \citep{mordasinialibert2012}. We assume that planetesimals (or their debris) always sink instantaneously to the core (discussion in Sect. \ref{sect:sinkingapproximation}). 

During gas runaway accretion when the planet is detached from the nebula  {\citep[so called detached or transition stage cf.][]{bodenheimerhubickyj2000}},  {and the gas accretion rate is given by the disk-limited value}, we consider two limiting cases for the shock structure: First cold accretion where the complete accretion shock luminosity caused by the infall of the gas $\lacc$ is radiated away and does not contribute to the luminosity $\lint$ that is used to calculate the internal structure of the planet. The accretion shock luminosity is approximately given as $G M \mdotxy/R$ for a planet of mass $M$ and radius $R$ accreting gas at a rate $\mdotxy$ ($G$ is the gravitational constant). Second we consider hot accretion without any radiative losses at the shock, so that the luminosity used to calculate the structure $\lint$ includes both $\lacc$ and the luminosity originating from the cooling and contraction of the material already incorporated in the planet. 

 {The method to calculate this contribution and the temporal evolution of a planet is based on total energy conservation as described in  \citet{mordasinialibert2012}. In this approach, the specific entropy of the newly accreted material $S_{\rm new}$ is not an explicit input to the calculations. But a posteriori, it can be estimated by comparing the change in entropy in the convective zone during a time interval $dt$, and the approximation that the entropy changes during this time only due to accretion of new gas, and not due to cooling of the material already in the planet. This holds approximately for cold ($S\lesssim9.5$ in units of $k_{\rm B}$ per baryon as everywhere in this paper) high-mass planets ($M\gtrsim3-5\mj$)  where the Kelvin-Helmholtz timescale is much longer than the accretion timescale  (see Sect. \ref{sect:self-amplifying}). Such cases have approximately horizontal entropy curves immediately after the end of accretion (bottom left panel of Fig. \ref{fig:tlrtnom}).  For entropies $S$ in the convective zones at time $t$ and $t+dt$, we then have $S(t+dt)\approx f_{\rm new}S_{\rm acc}+(1-f_{\rm new})S(t)$, where $f_{\rm new}$ is the mass fraction that has been accreted during $dt$, $(M(t+dt)-M(t))/M(t+dt)$. From these considerations, we find estimates (actually upper limits due to the neglect of cooling) for the difference in entropy between the newly accreted material and the material in the convective zone. One finds that the newly accreted material has an entropy that is lower by about 0.2 to 1.2  $k_{\rm B}$/baryon while a planets grows from 3 to 10 $\mj$, with smaller differences at higher masses, and initially larger differences for lower core masses. } 

\subsubsection{Boundary conditions}
{During the detached stage, the outer boundary conditions for the internal structure are the same as in \citet{mordasinialibert2012} }
\begin{align}
  P&=P_{\rm neb}+\frac{\dot{M}_{\rm XY,max}}{4 \pi R^{2}} v_{\rm ff}+\frac{2 g}{3 \kappa} & \tau&={\rm max}\left(\rho_{\rm neb}\kappa_{\rm neb}R, 2/3\right) \label{eq:pdetached} \\
    T_{\rm int}^{4}&=\frac{3 \tau L_{\rm int}}{8 \pi \sigma R^2} & T^{4}&=(1-A) T_{\rm neb}^{4}+T_{\rm int}^{4}. \label{eq:tdetached}
\end{align}

{The surface pressure $P$ thus contains the contribution from the nebula $P_{\rm neb}$ (which quickly becomes negligibly small relative to the other terms), from the ram pressure ($v_{\rm ff}$ is the free fall velocity), and from the photospheric pressure in the Eddington approximation ($g$ is the gravitational acceleration on the planet's surface, $\kappa$ the Rosseland mean opacity). The expression for the optical depth $\tau$ includes the effect of the material surrounding the  planet \citep{papaloizounelson2005}, but quickly becomes equal to $2/3$,  because the radius $R$ decreases rapidly by roughly two orders of magnitude after detachment. }

{The surface temperature  is given by the irradiation from the nebula with a temperature $T_{\rm neb}$ assuming a Jupiter-like albedo $A$ and the contribution due to the planet's own luminosity, $T_{\rm int}$, which quickly after detachment becomes clearly dominant. As mentioned, $L_{\rm int}$ does (does not) include $L_{\rm  acc}$ in the hot (cold) case. }

{These boundary conditions differ from the ones of  \citet{bodenheimerhubickyj2000} and \citetalias{marleyfortney2007} in two aspects: First, they include the additional term due to the photospheric pressure. This is strictly speaking not self-consistent with a fully spherically symmetric geometry of the infall, because the photospheric pressure is derived under the assumption that the gas above the $\tau=2/3$ surface is in hydrostatic equilibrium, which is not the case in such a geometry. On the other hand, in more complex geometries, the accreted material joins the planetary surface only in certain parts as suggested by 3D hydrodynamic simulations, namely through a circumplanetary disk \citep{tanigawaohtsuki2012,uribeklahr2013} near (but maybe not exactly at) the midplane and/or polar regions \citep{ayliffebate2012}. In the regions without much gas accretion, the photospheric pressure would then (approximately) apply \citep{hartmanncassen1997}. We  tested the impact of omitting the photospheric pressure on the post-formation entropy, finding rather small changes (see  Appendix \ref{sect:modelsettingsMcoreS0}).}

{The second difference in the boundary conditions relative to \citet{bodenheimerhubickyj2000}  is related to the temperature. We  simply assume that the material falling from the Hill sphere onto the planet is optically thin to the outgoing radiation. The model of  \citet{bodenheimerhubickyj2000} is more refined, since they solve the radiative transfer equation in the diffusion approximation using a density profile given by the free fall and a constant opacity. They find that this leads to good agreement with more detailed solutions of the radiative transfer equation. It is clear that the detailed structure of the temperature of the infalling gas in front of the shock (e.g., concerning the radiative precursor, or the distinction between the optically thick  or thin regime) is of high interested for more precise predictions about the planet's initial entropy.}

{To model this at least in 1D, it should be possible to use radiative hydrodynamic models similar to those of the accretion shock of the Larson's first core during star formation \citep[e.g.,][]{commerconaudit2011}. At this stage of star formation, it is sufficient to use the grey approximation in the radiative transfer calculations \citep{vaytetaudit2012} because one is dealing mostly with IR radiation. For the accretion shock on the (second) hydrostatic stellar core considered by \citet{stahlershu1980}, the temperatures become considerably higher, so that radiation over a wide frequency range (IR to X-rays) becomes important. As indicated by the calculations of \citet{stahlershu1980}, it is then necessary to consider a frequency-dependent treatment of the radiation.  We note that the shock temperature $T_{\rm shock}$  in the planetary case can in principle become quite high: for a supercritical shock \citep{commerconaudit2011}}
\beq
T_{\rm shock}=\left(\frac{1/2 \rho_{\rm ff} v_{\rm ff}^{3}}{\sigma_{\rm SB}}\right)^{1/4}
\eeq
{where $\rho_{\rm ff}=\dot{M}_{\rm XY}/(4 \pi R^{2} v_{\rm ff})$ and $\sigma_{\rm SB}$ is the Stefan-Boltzmann constant. For a 10 $\mj$ accreting at $10^{-2}\mearth/$yr  in the cold case (radius only about 1.4 $\rj$, see Table \ref{tab:res10mj}) $T_{\rm shock}\approx  7600$ K, leading to important contribution from the UV to the IR domain, potentially making a frequency dependent treatment necessary. }

{For the determination of the appropriate boundary conditions, the detailed description of the accretion shock itself is however not  yet sufficient. \citet{hartmanncassen1997} study cold accretion onto fully convective low-mass stars, a situation closely related to the one here, at least before D-ignition. They show that for accretion through a circumstellar disk, for large parts of the surface of the star, normal photospheric boundary conditions apply. This leads to a more efficient cooling compared to the spherical accretion geometry of  \citet{stahlershu1980}, meaning that the global accretion geometry also matters. For giant planets, we may expect that initially, when the Hill sphere radius of the planet is still smaller than the disk's vertical scale height, the accretion is roughly spherically symmetric, while later on, the accretion occurs through a disk, with the later phase being dominant for massive planets \citep{dangelolubow2008}. The gas in the circumplanetary disk would then eventually accrete into the planet via a  viscous boundary layer or potentially magnetospheric accretion columns as for stars \citep{lovelacecovey2011}.   }

Once the desired total mass of a planet is approached in a simulation, the gas accretion rate is shut down on a short timescale. During the evolutionary phase at constant mass we use as \citet{bodenheimerhubickyj2000} a simple gray atmosphere with solar composition opacities from \citet{freedmanmarley2008}. This means that our results for the long term cooling are less accurate than those of models using non-gray models \citep[e.g.,][]{burrowsmarley1997,baraffechabrier2003}.

\begin{table}
\caption{Settings for the formation and evolution calculation.}\label{tab:evocalcs}
\begin{center}
\begin{tabular}{lc}
\hline\hline
Quantity & Value \\ \hline
a [AU] & 5.2 \\
 $\sigmas0$ [g/cm$^{2}$]                         & 10, 12, 15, 18   \\                                              
$\dot{M}_{\rm XY,max}$ [$\mearth$/yr]& 0.01 \\
$T_{\rm neb}$ [K]           &150\\
$P_{\rm neb}$ [dyn/cm$^{2}$]           &   0.275     \\ 
Dust to gas ratio  &1/70 \\
Initial embryo mass [$\mearth$] & 0.1\\
Migration & not included\\
Disk evolution & not included\\
Planetesimal ejection & included\\
Core density & variable\\
Planetesimal size & 100 km \\
Fate of dissolved planetesimals & sink to core interface\\
Grain opacity red. factor $\fopa$ &0.003\\
Accretion shock luminosity & completely radiated \\ 
Atmosphere & gray approximation \\
Simulation duration & 20 Gyrs\\ \hline
\end{tabular}
\end{center}
\end{table}

\section{Combined formation and evolution simulations}\label{sect:simulations}
\begin{figure*}
\begin{minipage}{0.34\textwidth}
	      \centering
       \includegraphics[width=1\textwidth]{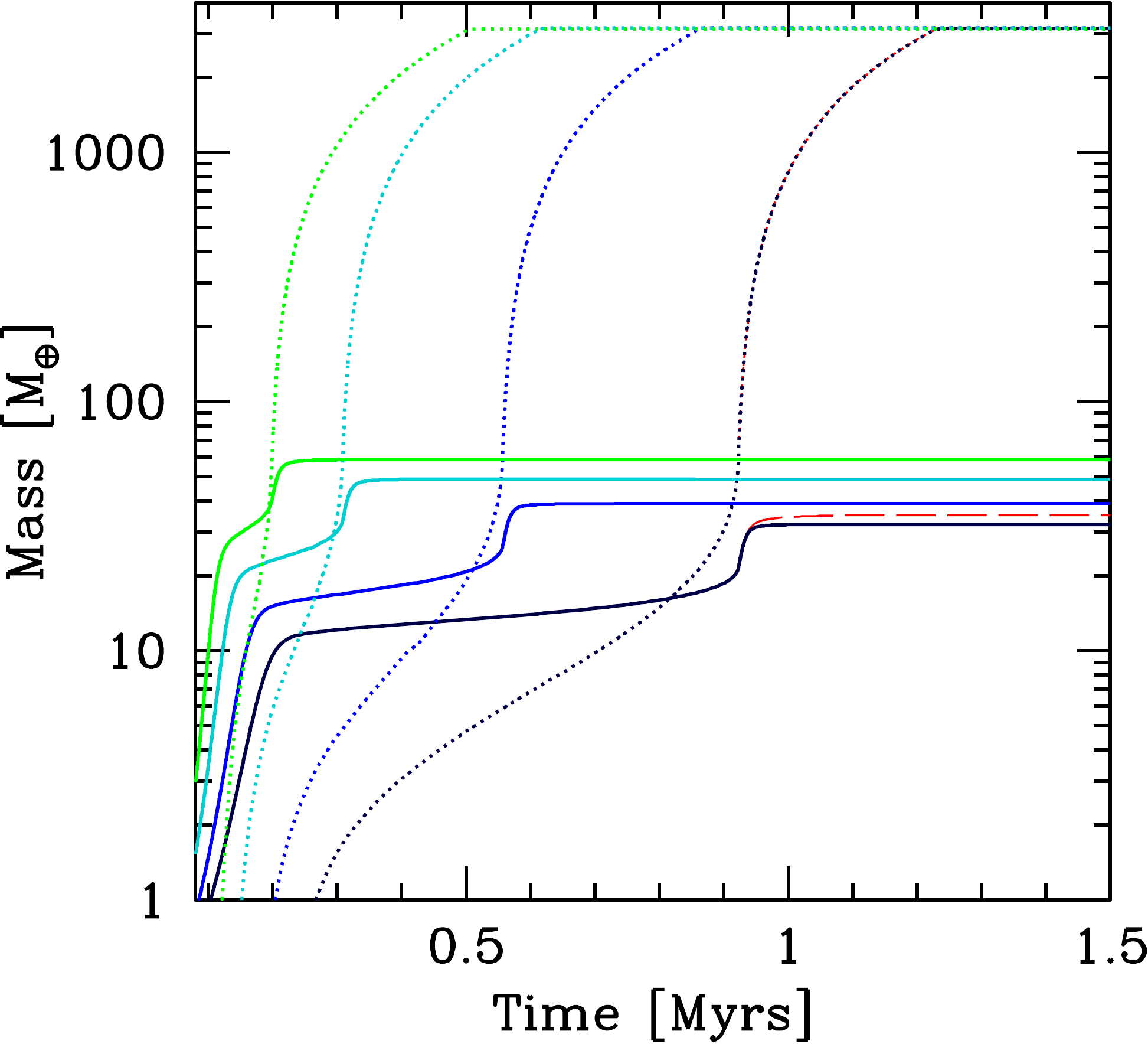}
                     \includegraphics[width=1\textwidth]{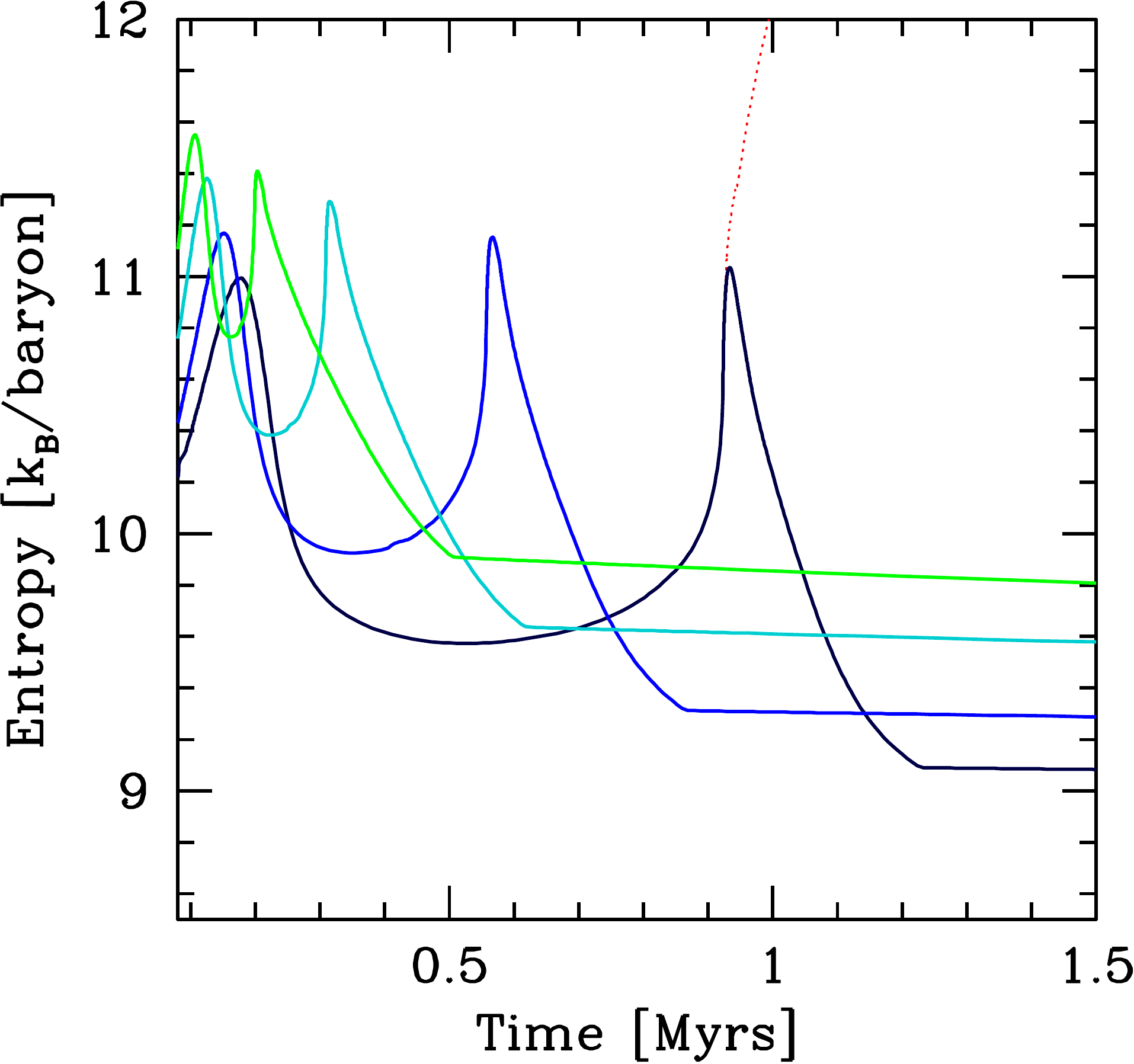}
     \end{minipage}
     \begin{minipage}{0.34\textwidth}
      \centering
                    \includegraphics[width=1\textwidth]{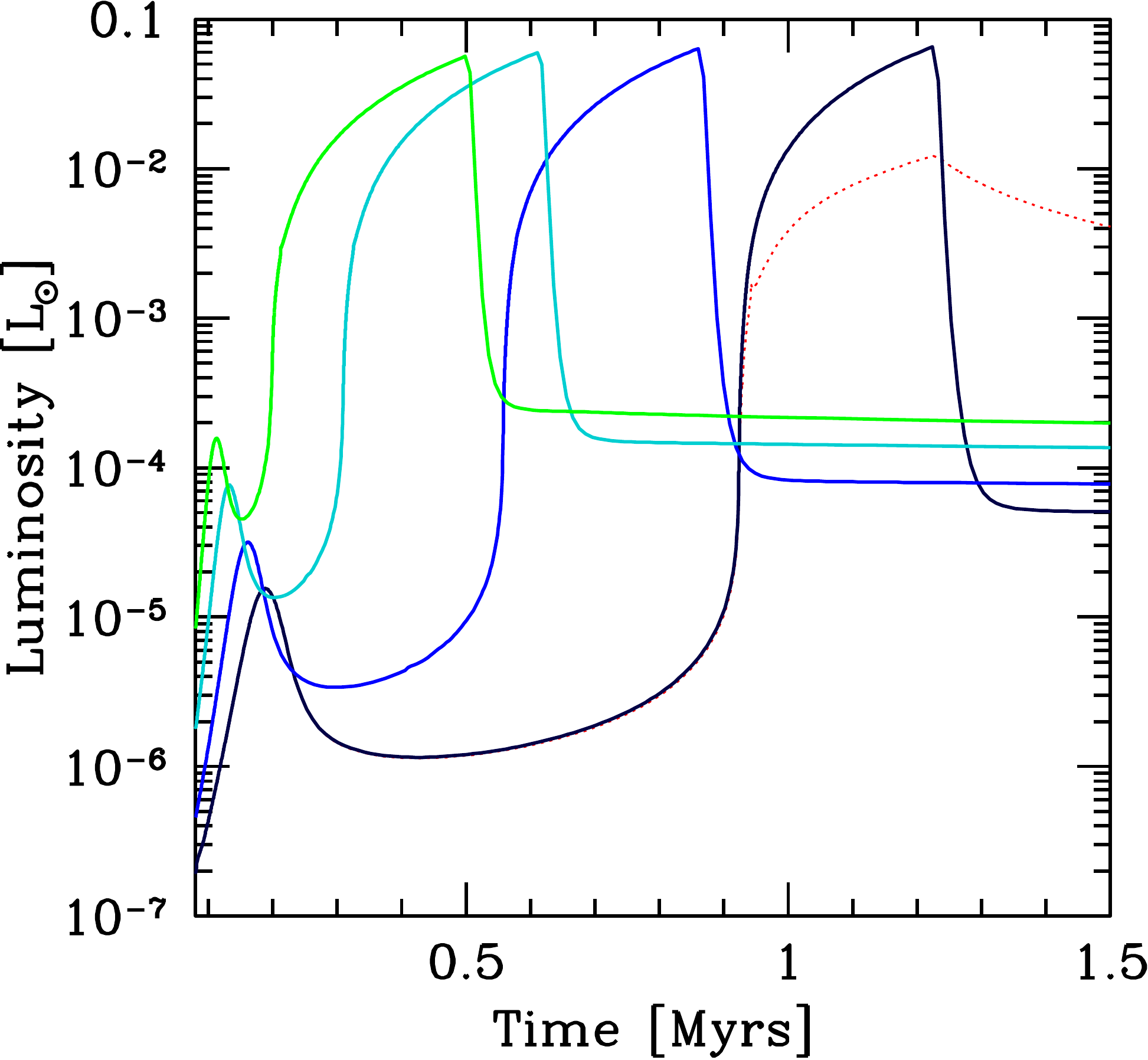}
       \includegraphics[width=1\textwidth]{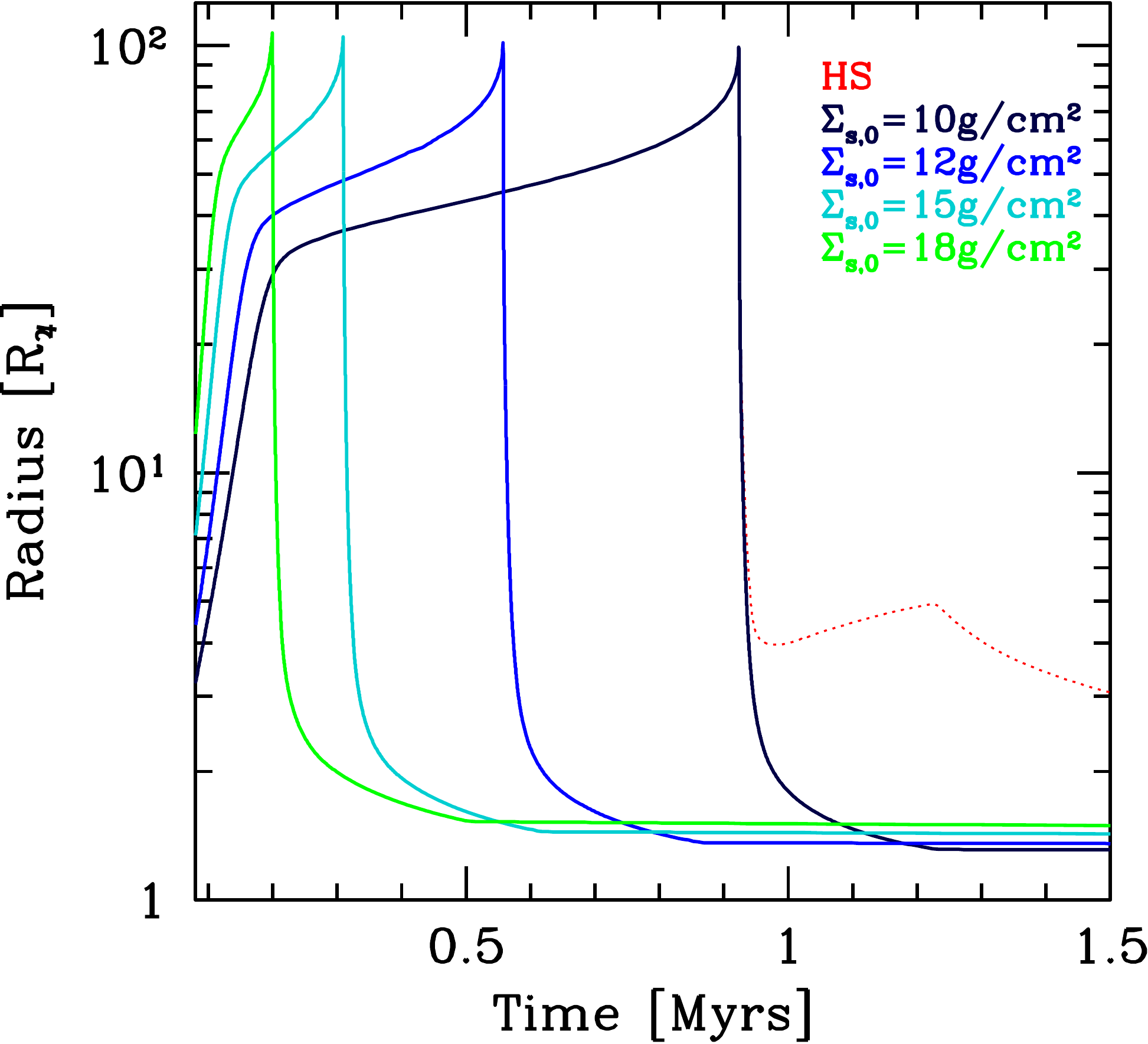}
     \end{minipage}
     \begin{minipage}{0.3\textwidth}  
     \vfill
     \hfill
                 \caption{Formation and early evolution of 10 $\mj$ planets forming at 5.2 AU.  The thin red line is for hot accretion (HS). The remaining four cases are all for cold accretion, but differ in the initial planetesimal surface density $\sigmas0$ as indicated in the plot. The top left panel shows the core (solid) and total mass (dotted lines). The top right panel shows the total luminosity. The bottom left and right panels are the specific entropy at the core-envelope boundary, and the outer radius,  respectively. Note how the luminosity, entropy and radius at the end of the formation phase increase with core mass for cold accretion.}
             	\end{minipage}
             \label{fig:tlrtnom} 
\end{figure*}

Using the model  described in the last section, we simulate the formation and evolution of planets of different total and core masses with the goal to determine observable post-formation properties like luminosity and radius. Before we discuss the post-formation properties, we show the  formation (and early evolution) of massive giant planets for different initial solid surface densities $\sigmas0$.

\subsection{Formation of a 10 $\mj$ planet for different $\sigmas0$}
Figure \ref{fig:tlrtnom} shows the formation and early evolution of five 10 $\mj$ planets. Four different initial solid surface densities of planetesimals  $\sigmas0$ are considered: 10 (nominal), 12, 15, and 18 g/cm$^{2}$. Four simulations are calculated under the cold accretion assumption, while an additional fifth comparison simulation  also has $\sigmas0$=10 g/cm$^{2}$, but assumes hot accretion. The simulations are otherwise identical. The most important results of the simulations are listed in Table \ref{tab:res10mj}.

\subsubsection{Mass}\label{sect:simsmass}
The top left panel shows the total mass $M$ and core mass $\mcore$ as a function of time. The higher $\sigmas0$, the shorter the formation time $\tform$ until the final mass of 12 $\mj$ is reached, and the higher the final core mass, as expected \citep[e.g.,][]{pollackhubickyj1996}. The planets in this simulation continue  to  accrete planetesimals after crossover and in the runaway phase. This leads to a two step accretion of solids: first in phase I\footnote{{The phases I, II, and III refer to the main phases that are seen in classical in situ formation simulations of Jupiter, as defined in \citet{pollackhubickyj1996}. During phase I the embryo accretes rapidly all planetesimals in the feeding zone, until the core reaches the isolation mass. The envelope mass is very small due to the high core luminosity. During phase II, a slow growth of the core and the gaseous envelope occurs where the planet has to accrete gas in order to expand its solid feeding zone. The accretion rate of gas is higher than the one of planetesimals. Phase III begins at the crossover point, i.e., the moment in time when the core and envelope mass become equal. Afterwards, a rapid increase of the gas accretion rate occurs, leading to runaway gas accretion. Note that it is  far from clear that these phases actually occur: Phase I is only possible if the random velocities of the planetesimals stay small, which is clearly questionable for large planetesimals \citep[e.g.,][]{idamakino1993}. Phase II can be suppressed by orbital migration \citep{alibertmordasini2004} or the opening of a gap in the planetesimal disk \citep{shiraishiida2008}.}}, and then again at  and after crossover due to the rapid extension of the feeding zone that is proportional to the Hill sphere radius $\rhill$. The final mass is then limited by the combined effects of planetesimal ejection and emptying of the feeding zone. During phase II and also during runaway, if the planet accretes all planetesimals in the feeding zone, it holds that $\mcore=(M \miso^{2})^{1/3}$ \citep{pollackhubickyj1996} where $\miso$ is the isolation mass \citep{lissauer1993}. Since $\miso\propto \sigmas0^{3/2}$, this means that for a fixed $M$ the final core mass is proportional to $\sigmas0$, which holds very well in the simulations (Tab. \ref{tab:res10mj}). The final core masses are between 32 and 59 $\mearth$, i.e., they differ by a factor 1.8 as  $\sigmas0$. {Regarding the mass,} the hot accretion case is basically the same as the corresponding cold case, except for a slightly higher final core mass. It is the consequence of a larger capture radius due to an overall larger planet radius. 

\subsubsection{Luminosity}\label{sect:lumi4sims}
The top  right panel shows the total luminosity, i.e., it includes for the cold accretion case both $\lint$ and $\lacc$. One sees the characteristic shape \citep[e.g.,][]{bodenheimerhubickyj2000} with a first lower maximum during phase I that is due to the accretion of the core, then a low luminosity during phase II, and finally the gas runaway accretion phase where the planets are very bright. For cold accretion, the luminosity during this phase is increasingly dominated by $\lacc$ that is a factor $\sim$1000 bigger than $\lint$ at the moment the gas accretion rate is artificially reduced because a mass of 10 $\mj$ is approached. The overall largest luminosity is reached just before the reduction of the gas accretion begins ($L_{\rm rga}$). While difficult to see on the logarithmic scale, the $L_{\rm rga}$ for the planet with the most massive core is actually about 15\% lower than in the simulation with the lowest core mass. This difference is relevant, as will become clear below. During the reduction of $\mdotxy$, the luminosity in the cold accretion simulations falls sharply since the contribution of $\lacc$ vanishes. Once the final mass has been reached, the luminosity takes the value of the intrinsic luminosity at the end of formation respectively at the beginning of evolution, $L_{\rm 0}$. The central result of this paper is now that this $L_{\rm 0}$ is not identical in the four cold accretion simulations. Instead, it is a strong function of the core mass: $L_{0}$ is $5.1\times10^{-5} \lsun$ for $\mcore=32\mearth$, but $2.4\times10^{-4}\lsun$ for $\mcore=59\mearth$. The increase of $\mcore$ (and $\sigmas0$) by a factor 1.8  has thus caused an increase of $L_{0}$ by a factor 4.7, which corresponds for this interval in core mass to a power-law dependence roughly like $\mcore^{2.6}$ {(see Appendix \ref{sect:l0mcore} for a discussion of the power-law approximation.)} The origin of this effect is discussed in Section \ref{sect:self-amplifying}. As we will see further down in Sect. \ref{sect:luminosities}, the effect is strong enough to have potentially observable consequences. 

\subsubsection{Entropy}
The bottom left panel shows the entropy $S$ at the core-envelope interface. As expected from the luminosity curve, there are two maxima, one during phase I, and a second one shortly after detachment ($S_{\rm max}$, cf. Table \ref{tab:res10mj}).  The planets detach from the disk at a mass of about 100 $\mearth$ (slightly increasing with $\sigmas0$),  and then contract initially very rapidly. The maximal entropy is reached when the outer radius has decreased to about 6-10 $\rj$ {(Jovian radii)}, and the mass is around 150 to 200 $\mearth$. After this maximum that is higher for more massive cores, the entropy steadily decreases due to the addition of low-entropy material for cold accretion. Note that the four entropy curves diverge during this decrease, i.e., the relative differences in $S$ between the four simulation increases. At the end of formation, there is a difference in the entropies $S_{0}$ of  0.81. For hot accretion, the entropy in contrast increases during the entire runaway phase to reach a maximum value $S_{\rm max}$=$S_{0}$=13.31  just before accretion is terminated (outside of the figure).

\subsubsection{Radius}\label{sect:simsradius}
The bottom right panel shows the outer radius. The basic shape of the curves follows the pattern found by \citet{bodenheimerhubickyj2000} with a large radius during the attached phase ($\sim \rhill$), then the rapid collapse, and finally the slower contraction once the interior becomes sufficiently degenerate. We see that the planet with the most massive core has the largest radius $R_{\rm 0}$ at the end of the formation epoch. Its higher entropy leading to a larger radius \citepalias[e.g.,][]{spiegelburrows2012} thus dominates at least initially over the effect that a higher heavy element fraction reduces the radius \cite[e.g.,][]{fortneymarley2007a,baraffechabrier2008a}. As the planets cool, the radii decrease and the memory of the initially higher entropy gets lost, so that at about 120 Myrs, the radius of the planet with the most massive core and the one of the planet with the lowest core mass cross over, and afterwards, the planet with the more massive core has a smaller radius, as it is normally the case (see Sect. \ref{sect:radii}). 

For cold accretion, the planets' radii decrease all the time during runaway accretion while they grow in mass. For hot accretion, the planet in contrast first collapses rapidly to a radius of about 4 $\rj$ (when $M$$\approx$2$\mj$) and then re-inflates to $R_{\rm 0}$=4.94$\rj$ at the end of formation due to the addition of high-entropy material.  This $R_{\rm 0}$ is similar to the ``initial'' radius  assumed in evolutionary hot start simulations like, e.g., \citetalias{spiegelburrows2012} who use $R_{\rm 0}$=4.15$\rj$ (the ``initial'' radius is of course ill defined to the extremely short $\tKH$). This agreement indicates  that our simplification for the solution of the internal structure equations (constant luminosity as a function of radius) does not lead to strong departures from the typical hot start scenario, despite the  complications (absence of a  deep radiative zone) the $\partial L/\partial r=0$ approximation could potentially have for hot accretion  \citep[see][]{mordasinialibert2012a}. 

\begin{table}
\caption{Results for 10 $\mj$. For $L_{\rm 0}$ and $L_{\rm rga}$ the number in parentheses is the exponent $p$ in $10^{p}$. }\label{tab:res10mj}
\begin{center}
\begin{tabular}{lccccc}
\hline\hline
$\sigmas0$                   &10 &  12 & 15 & 18 & HS\\\hline 
$\miso$ [$\mearth$]    &11.5 & 15.1 & 21.1& 27.8 & 11.5 \\   
$\tform$  [Myrs]          &  1.25 & 0.87 & 0.62 & 0.51 &  1.25\\
$\mcore$ [$\mearth$]	 & 32.1 & 38.9 & 48.9 & 58.6 & 35.1 \\
$\s0$ [$k_{\rm B}$/ba.]&  9.09 & 9.31 & 9.63 & 9.90 & 13.40\\
$R_{\rm 0}$ [$\rj$]       & 1.31 & 1.36 & 1.44 & 1.52 & 4.94   \\
$L_{\rm 0}$ [$L_{\odot}$]  & 5.1(-5) & 8.1(-5) & 1.5(-4) &2.4(-4)& 1.2(-2)\\
$S_{\rm max}$ [$k_{\rm B}$/ba.]	 & 11.03  & 11.15 & 11.29 & 11.41 & 13.40\\
$L_{\rm rga}$ [$L_{\odot}$] &6.5(-2)  &6.3(-2)  & 6.0(-2) &5.6(-2)& 1.2(-2)\\ \hline
\end{tabular}
\end{center}
\end{table}

\subsection{Self-amplifying process}\label{sect:self-amplifying} 
One may wonder how a change in the core mass by a small fraction of the planet's total mass (27$\mearth$ or just 0.8\% of the total mass) can significantly affect the luminosity of 10 $\mj$ planet. The reason is a self-amplifying process during  gas runaway accretion, combined with the fact that for a higher $\sigmas0$, the entropy is already higher when runaway gas accretion starts. The latter is due to the fact that for a higher $\sigmas0$, a more massive core (and envelope) forms on a shorter timescale, which means that the luminosity must be higher and so the entropy.  This leads to a $S_{\rm max}$ that is a factor 1.03 larger in the simulation with the most massive core relative to the nominal case. A consequence of the higher entropy at detachment is that the outer radius of the high $\mcore$ planet is larger for a given mass. This means that the gas that is accreted during runaway falls less deep in the potential well of the planet, so the shock is weaker and $\lacc\approx G M \mdotxy/R$ is smaller. Less liberated potential energy is thus  radiated, instead the gas is incorporated at a higher entropy into the planet. The higher entropy causes a larger radius, which again means a weaker shock and so on, i.e., there is a self-amplifying process. This means that the relative difference in entropy between the different $\mcore$ cases continuously increases in the detached phase. Indeed, the ratio of the entropy $\s0$ in the planet with the most massive core relative to the nominal case has increased to 1.09 at the end of formation.  Due to the strong sensitivity of the luminosity on $S$ \citep[e.g.,][]{marleaucumming2013} such factors matter. 

At the same mass, a planet with a higher entropy would usually contract  faster because the Kelvin-Helmholtz timescale $\tKH$ is shorter. This would stop the self-amplifying process. But for the assumed gas accretion rate in runaway of $\dot{M}_{\rm XY,max}=10^{-2} \mearth$/yr, the accretion timescale $\tacc= M/\dot{M}_{\rm XY,max}$ is most of the time shorter than $\tKH$ (calculated as $|E_{\rm tot}|/\lint\sim G M^{2}/ (R \lint)$ where $E_{\rm tot}$ is the sum of the gravitational and thermal energy in the planet). Early after detachment, approximately at the same time when $S_{\rm max}$ occurs, $\tKH/\tacc$ first falls slightly below unity, and the radii of the planets indeed start to converge. But the radius of the planet with the more massive core remains still larger, and with increasing mass, the ratio $\tKH/\tacc\propto M/(R \lint)$ increases ($\lint$ and $R$ decrease during the runaway accretion phase for $M\gtrsim300\mearth$ for cold accretion). Just before the gas accretion is ramped down, $\tKH/\tacc\approx600$ and 100, for $\mcore=32$ and 59 $\mearth$, respectively. This means that the planet with the higher core mass cannot sufficiently rapidly contract to stop the self-amplification. Instead, the initially relatively small entropy difference gets amplified, with the described consequences for the post-formation luminosity.

This mechanism also explains why $L_{\rm rga}$ is 15\% smaller in the simulation with the largest core mass and cold accretion than in the nominal case. For hot accretion $L_{\rm rga}$ (that is here the same as $L_{0}$) is even a factor 5.4 smaller than in the nominal cold accretion simulation. The smaller the luminosity during gas runaway accretion, the larger the luminosity during evolution, because the planet stores via a higher entropy the part of the gravitational potential energy  that is not radiated away during formation. While this fundamental mechanism is relatively general, it is clear that the quantitative consequences (specific values of $S_{0}$) depend on assumptions like the gas accretion rate during runaway \citep[that is certainly not constant, cf.][]{lissauerhubickyj2009,bodenheimerdangelo2013} or the infall geometry that is certainly not spherically symmetric \citep[e.g.,][]{ayliffebate2009} {(see also Appendix \ref{sect:modelsettingsMcoreS0})}.

\subsection{Sinking approximation}\label{sect:sinkingapproximation}
An important assumption necessary for the mechanism to work efficiently is that the planetesimals or their debris sink to the core (or at least deep into the planet) on a short timescale ($\ll\tacc)$, so that they liberate a significant amount of potential energy to heat the planet during the formation phase. Especially for the planetesimals accreted after crossover (i.e., during the second phase where the core grows significantly, see Sect. \ref{sect:simsmass}) it is unclear whether the sinking approximation holds. This is because the gaseous envelope at this moment is already hot and massive (several 10 $\mearth$) while simulations of the impacts of planetesimals \citep{podolakpollack1988,mordasinialibert2006} into protoplanetary atmospheres rather indicate that 100 km planetesimals can only directly penetrate through envelopes with masses of a few $\mearth$ \citep{mordasinialibert2006}. In more massive envelopes, the planetesimals get fragmented and then vaporized by shock wave radiation during the impact. For this material to also contribute efficiently  to the heating of the planet, the planetesimal debris has to sink after the impact. \citet{iaroslavitzpodolak2007} studied this by always comparing the partial and the saturation vapor pressure of the high-Z material in all layers during a formation simulation of \citet{hubickyjbodenheimer2005}  so that they can keep track where the high-Z material ends up. They find that even if much high-Z material does not reach the core in solid form directly, it is still concentrated (as vapor) in the envelope layers closely above it. They therefore conclude that the  gravitational energy release (which is crucial in this study) is similar as in the  sinking approximation we use here. The study of \citet{iaroslavitzpodolak2007} is simplified in a number of aspects and only follows the evolution of the debris up to the crossover mass, when core and envelope mass are equal. The fate of planetesimal material accreted after this point is thus less clear.  For this material to sink also, it would maybe be necessary that an equivalent mechanism occurs as for the heating of giant planets by He-separation \citep[e.g.,][]{stevensonsalpeter1977}: the metals would need to undergo de-mixing from the H/He \citep[e.g.,][]{wilsonmilitzer2012}, to nucleate and to grow into sufficiently large droplets in order to settle despite convention, and this on a sufficiently short timescale \citep{liagnor2010}. Whether this is possible  needs to be checked in future work.

\begin{figure}
\begin{center}
\includegraphics[width=0.95\columnwidth]{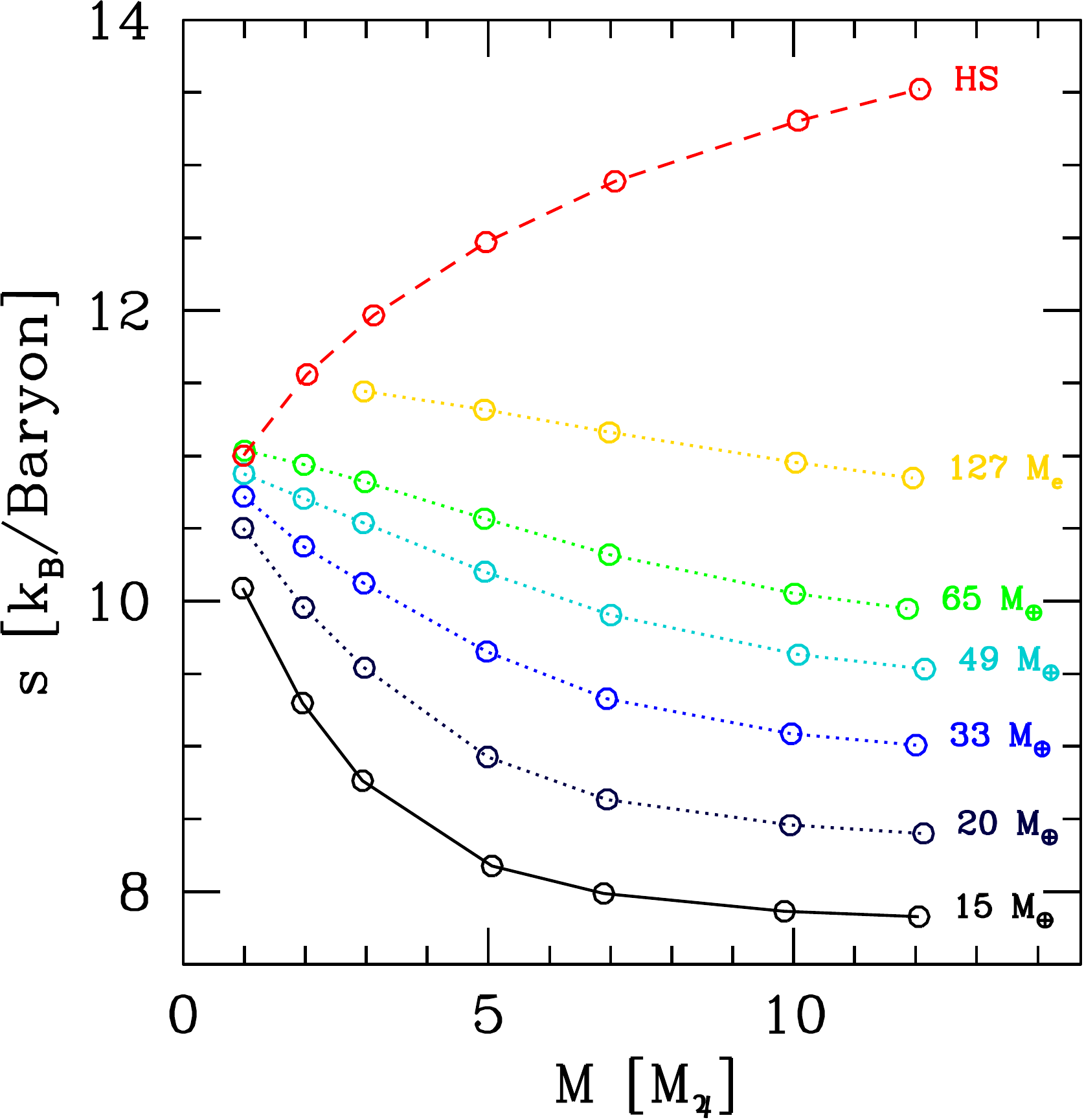}
\caption{Post-formation entropy as function of total mass and for six different core masses indicated in the plot for cold accretion. The red dashed line labelled HS is for hot accretion.}
\label{fig:fork}
\end{center}
\end{figure}

In order to test if it is important at which moment in time the planetesimals are accreted (during phase I or also after crossover), we  conducted simulations where the accretion of planetesimals is artificially shut down once crossover is reached, as it seems to be the case in the simulations of \citetalias{marleyfortney2007}. At this moment the core mass is approximately equal to $\sqrt{2}\miso$ \citep{pollackhubickyj1996}. We  then adjusted the initial surface density of planetesimals in a way that this mass is the same as the final core mass in the normal simulations where the planets continue to accrete planetesimals also after crossover. We find that the resulting post-formation luminosities are very similar. This means that for the core mass to influence the luminosity after formation, it is not necessary that the massive core forms during the runaway phase when it is, as explained, potentially more difficult for the solids to penetrate deep enough. It is sufficient that a massive core forms at all, especially during phase I and before crossover, when it seems that  the sinking approximation applies at least in an approximate way. But it is clear that to reach the same core mass without accretion during runaway, a larger $\sigmas0$  (approximately by a factor 1.5 to 2) is needed.

Note that in particular in the absence of sinking, a sufficiently high energy input due to impacts can lead to the formation of a deep radiative zone \citep{liagnor2010}, a process similar to the formation of a radiative zone during hot accretion onto an initially fully convective star \citep{mercer-smithcameron1984,prialniklivio1985,hartmanncassen1997}. For sufficiently high accretion rates, the stars inflate, which could in principle also lead to a weaker accretion shock. This will also be addressed in future work.

\section{Post-formation entropy as function of total and core mass}\label{sect:entropy}
We have conducted combined formation and evolution simulations as in the last section but for total final masses between 1 and 12 $\mj$, and $\sigmas0$ leading to final core masses between approximately 15 and 127 $\mearth$. Figure \ref{fig:fork} shows the resulting specific entropies at the core-envelope boundary. The entropies $\s0$ are shown immediately after the moment when a planet has reached its final mass, i.e.,  when the evolutionary phase begins.

One sees the characteristic ``entropy tuning-fork'' \citepalias[][\citealt{marleaucumming2013}]{marleyfortney2007,spiegelburrows2012}. For cold accretion, the entropy decreases with increasing mass since an increasingly large fraction of the planet's mass is accreted through the entropy reducing shock, while for hot accretion, it increases because an increasing large amount of liberated potential energy is stored in the planets. The new result is that the level of the cold branch of the fork increases with increasing core mass. For massive planets, the entropy is about 3 $k_{\rm B}$/baryon larger for the most massive core we have considered (127 $\mearth$) than for the lowest value of 15 $\mearth$. At a total mass of 12 $\mj$, this is more than half of the difference between the 15 $\mearth$ core cold case and the hot accretion case, i.e., this is a large difference. We thus find that the mass of the core leads to a wide continuum of intermediate warm starts \citepalias{spiegelburrows2012} for core accretion planets. While \citetalias{spiegelburrows2012} envisioned these warm starts to be the consequence of  intermediate radiative efficiencies of the accretion shock (which is a perfectly possible alternative explanation) we find that even for a fully efficient shock, there is a large range of post-formation entropies  if planets can have high (but not exorbitant) core masses. 

It is interesting to compare our results with those of \citetalias{marleyfortney2007}. One has to take into account that we show the entropy at $t-t_{\rm form}$=0 Myrs, whereas \citetalias{marleyfortney2007} show it after 1 Myr of evolution which matters for the low-mass planets because of their short KH timescale. We find that for cold accretion, a 1$\mj$ planet has at 1 Myr an entropy of 9.30 and 9.36 for core masses of 15 and 20 $\mearth$, respectively, while \citetalias{marleyfortney2007} found 9.25 for a core mass of 17 $\mearth$. The 10 $\mj$ planet in our simulations has at 1 Myr an entropy of 7.86 and 8.45  for core masses of 15 and 20 $\mearth$ (virtually identical as at  the beginning), which brackets the  8.25  found by \citetalias{marleyfortney2007} for a 19 $\mearth$ core. There is thus  good agreement between the two completely independent codes despite the multitude of assumptions that enter into them. For hot accretion, the differences are somewhat larger, as our planets have at 1 Myr entropies that are 0.2 to 0.4  lower. But this is not surprising, because \citetalias{marleyfortney2007} uses arbitrary initial entropies for the hot case, which still matter after 1 Myr of evolution. 

{Regarding the  absolute values of the post-formation entropy, we need  however to keep in mind that both codes are 1D. Multidimensional hydrodynamic simulations point in contrast to complex accretion geometries \citep{tanigawaohtsuki2012,ayliffebate2012,uribeklahr2013}. This means that the quantitative results found here, and in particular the specific core mass - initial entropy mapping should not be taken as final results. Rather, they are indicators of possible important mechanism and correlations. They should be reassessed and re-quantified as soon as multidimensional, (magneto-)radiation hydrodynamic simulations with sufficiently high resolution and a realistic equation of state can directly predict the initial entropy of planets. In Appendix \ref{sect:modelsettingsMcoreS0} we quantify the impact of some model settings on the initial entropy within our 1D model framework.}

\begin{figure*}[tb]
\begin{minipage}{0.48\textwidth}
	      \centering
       \includegraphics[width=1.0\textwidth]{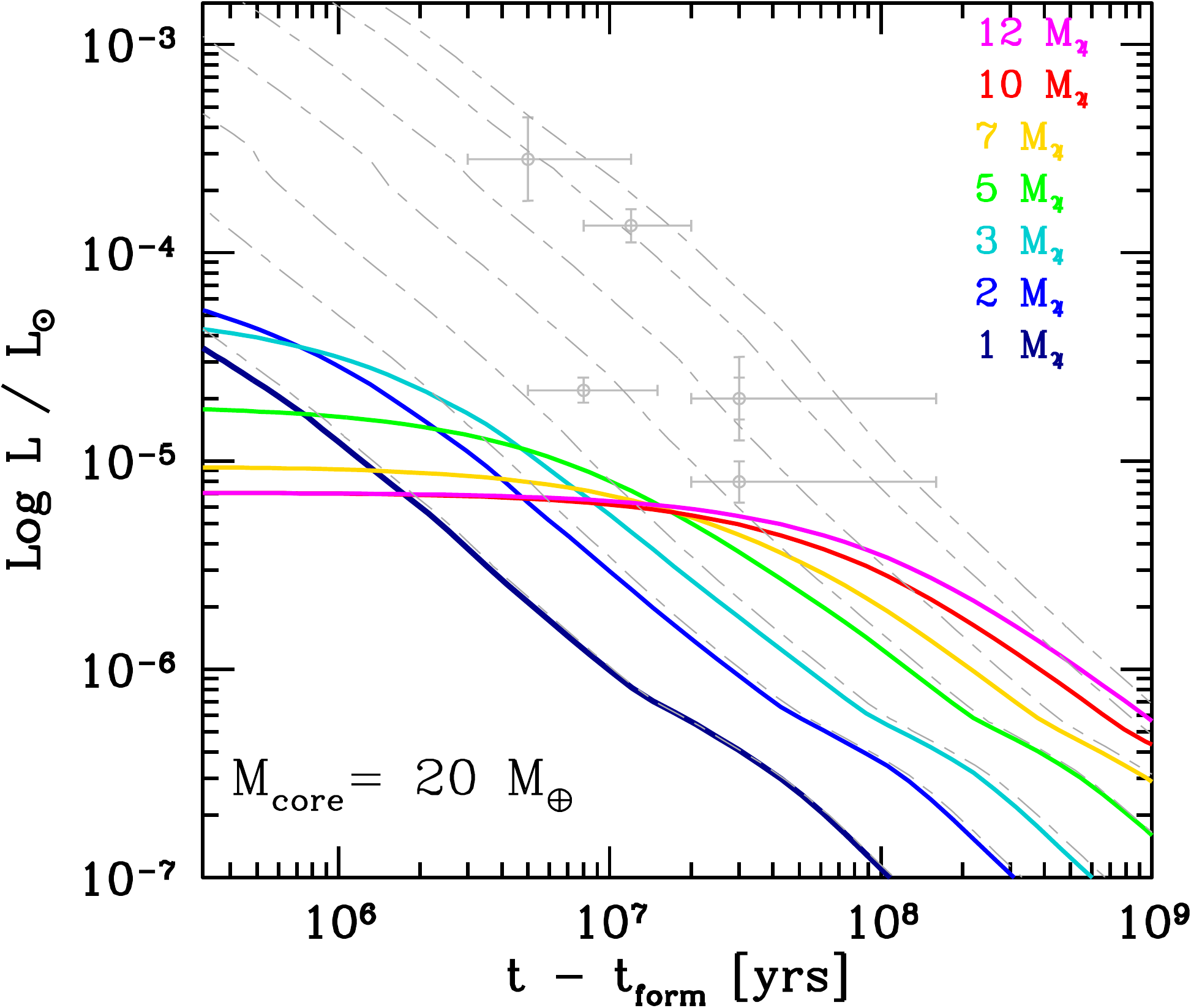}
              \includegraphics[width=1.0\textwidth]{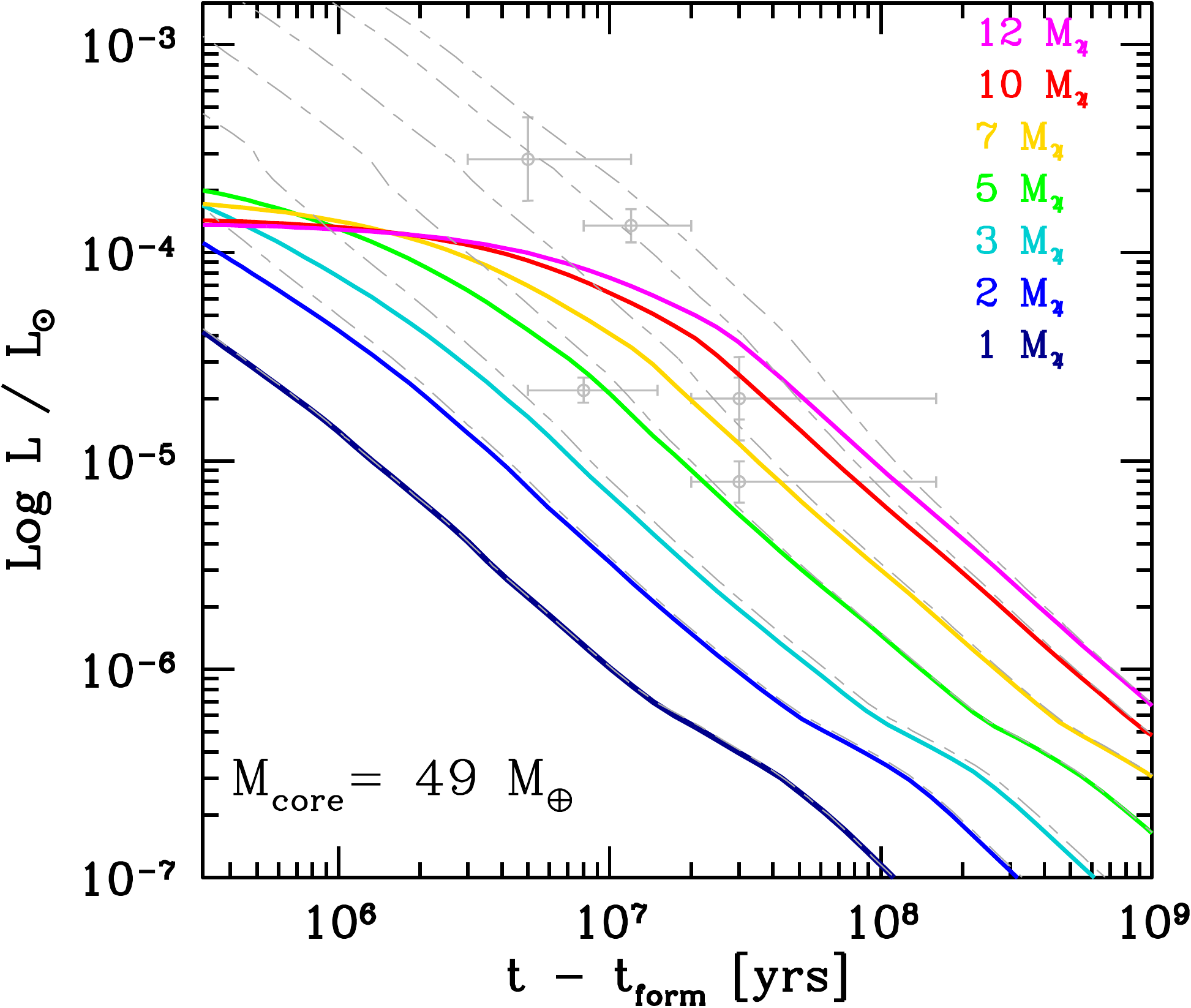}
     \end{minipage}\hfill
     \begin{minipage}{0.48\textwidth}
      \centering
       \includegraphics[width=1.0\textwidth]{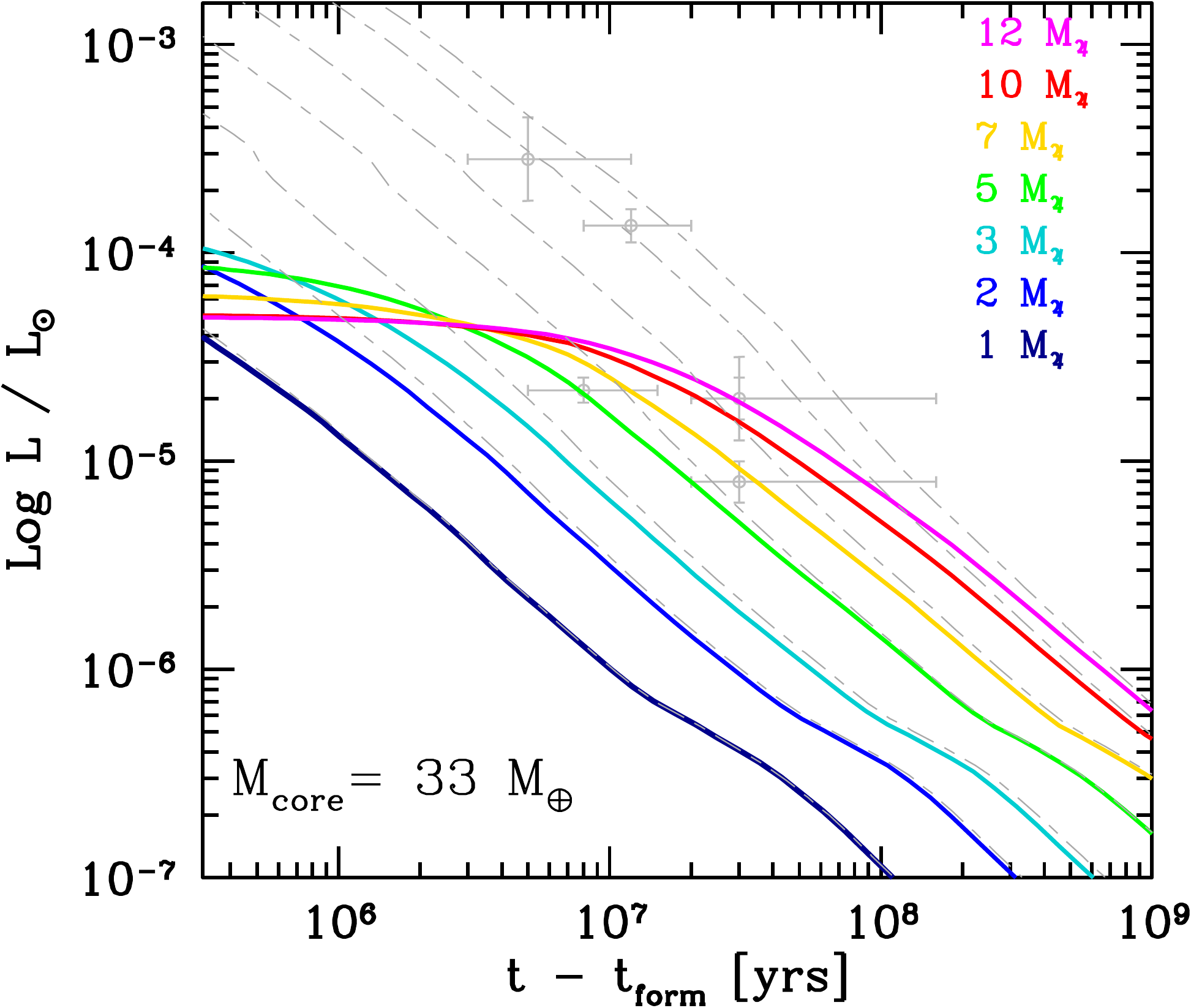}
              \includegraphics[width=1.0\textwidth]{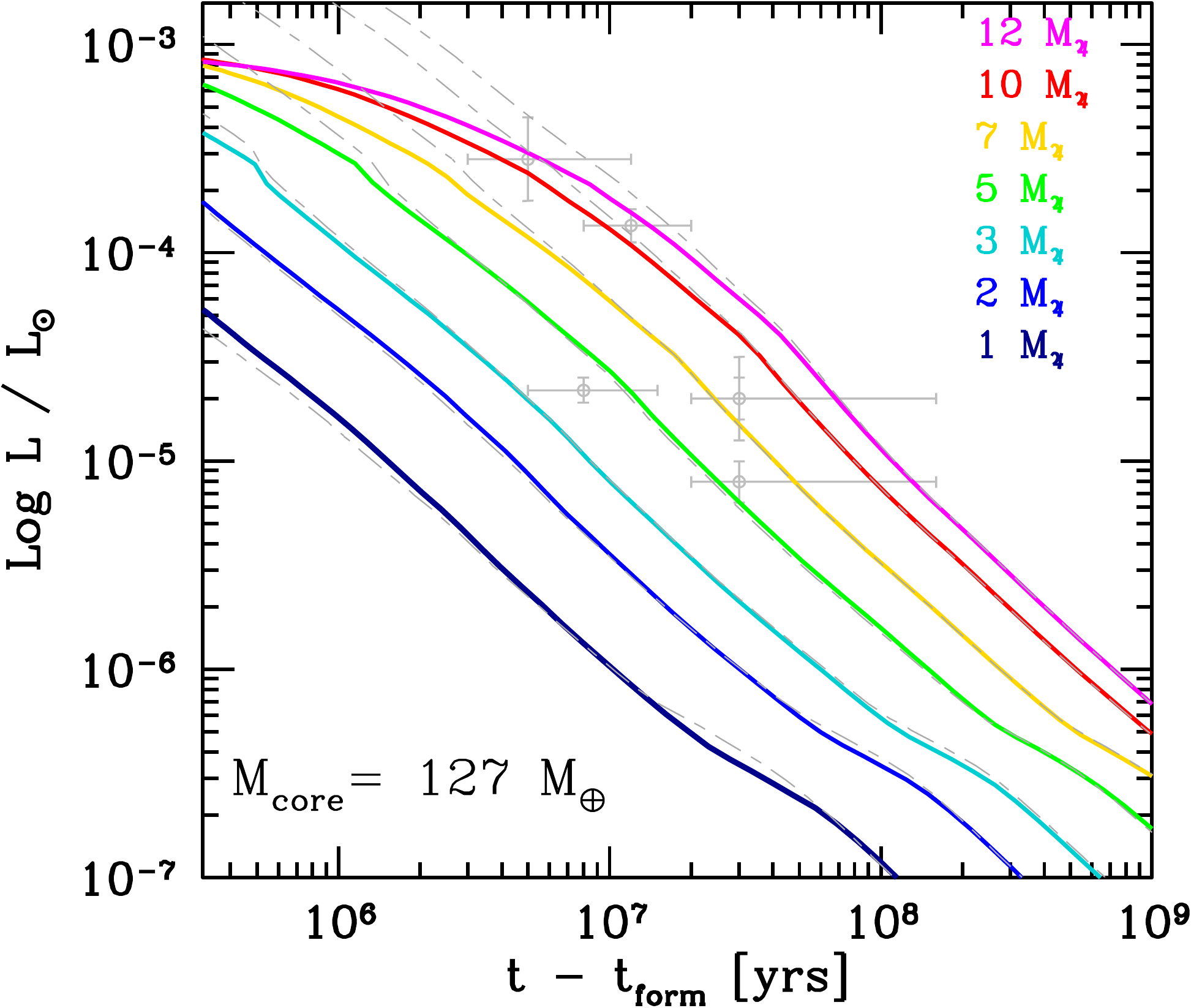}
     \end{minipage}
        \caption{Luminosity as a function of time after formation for planets with total masses of 1, 2, 3, 5, 7, 10, and 12 $\mj$ and core masses of 20, 33, 49, and 127 $\mearth$ as indicated in the panels. The colored solid lines assume cold accretion. The dashed-dotted gray lines assume hot accretion. The points with error bars are young giant planets (from top to bottom: 1RXS1609 b,  Beta Pic b,  2M1207 b,  HR8799c,d,e, HR8799 b).}\label{fig:mammamia} 
\end{figure*}

We finally note that for hot accretion, the mass of the core is not important: for an increase of the core mass from 35 to 132 $\mearth$ the post-formation entropy of a 10 $\mj$ planet grows by just 0.1, more than 20 times less than in the cold case. This is expected, because the accretion of gas represents for a radiatively inefficient shock a much stronger heating source than the accretion even of a massive core, and the self-amplifying mechanism described above can neither operate.  
  
\section{Luminosities during evolution}\label{sect:luminosities}
Figure \ref{fig:mammamia}  shows the luminosity of giant planets for different total  and core masses as a function of time after the final mass has been reached. The plot shows the results for cold accretion with four different core masses as well as for hot accretion. It additionally depicts the measured luminosities of seven young  giant planets. Note that the total and core masses of the model planets can vary slightly around the values indicated in the plots, since the masses are a product of the formation phase.

The basic result is the same as in earlier studies \citepalias{marleyfortney2007}: cold accretion leads to fainter planets, and the difference is larger and remains longer for more massive planets. This is due to the larger differences in post-formation entropies, and the longer Kelvin-Helmholtz timescale of more massive planets. The plot  shows that quantitatively, however, the magnitude of the difference between cold and hot accretion  strongly depends on the mass of the core.  The higher the core mass, the closer the cold and hot accretion cooling tracks. The impact of $\mcore$ increases with total mass. For a 10 $\mj$ planet at 1 Myr, for example, the luminosity with the most massive core is a factor $\sim$100 larger compared to the lowest core mass, and therefore only a factor 2 to 3 lower than for hot accretion. 

The much higher post-formation entropy of high $\mcore$ planets also strongly reduces the time until cold and hot accretion cooling curves converge, due to the exponential dependency of the KH timescale on $S$ \citepalias[e.g.,][]{marleyfortney2007}. At a core mass of $20$$\mearth$, it takes of order $10^{9}$ yrs until the luminosities of a 10$\mj$ planet become similar for cold and hot accretion, as already found by \citetalias{marleyfortney2007}, whereas at $\mcore=$127$\mearth$, it just takes $\sim$$10^{7}$ yrs.

The plot also shows the luminosities of  1RXS1609 b \citep{lafrenierejay2008},  $\beta$ Pic b \citep{lagrangebonnefoy2010},  2M1207 b \citep{chauvinlagrange2004},  and the HR 8799 planets \citep{maroismacintosh2008} as calculated by \citet{bonnefoyboccaletti2013}. While we do not make any claim that our models are specifically applicable to any particular object, the comparison of these objects with the cold accretion cooling tracks makes the importance of $\mcore$ clear. At $\mcore=20\mearth$, cold accretion is not consistent with any observation because the planets are too faint (as previously noted in several works), while for a core roughly 6 times more massive, all observed luminosities can be reproduced, and the mass estimates are virtually identical as for hot accretion for the less bright planets, and less than 2$\mj$ higher for the brighter ones.

The dependency on $\mcore$ makes the conversion of luminosity into mass for cold accretion  difficult: For HR 8799 b, for example, one deduces a mass clearly exceeding 12 $\mj$ for $\mcore=20$$\mearth$, but masses of $\sim$12,  9, and 7 $\mj$ for $\mcore$=33,  49, and  127 $\mearth$, the latter value being identical with hot accretion.

\section{Radii during evolution}\label{sect:radii}
\begin{figure*}[htb]
\begin{minipage}{0.33\textwidth}
	      \centering
       \includegraphics[width=1.0\textwidth]{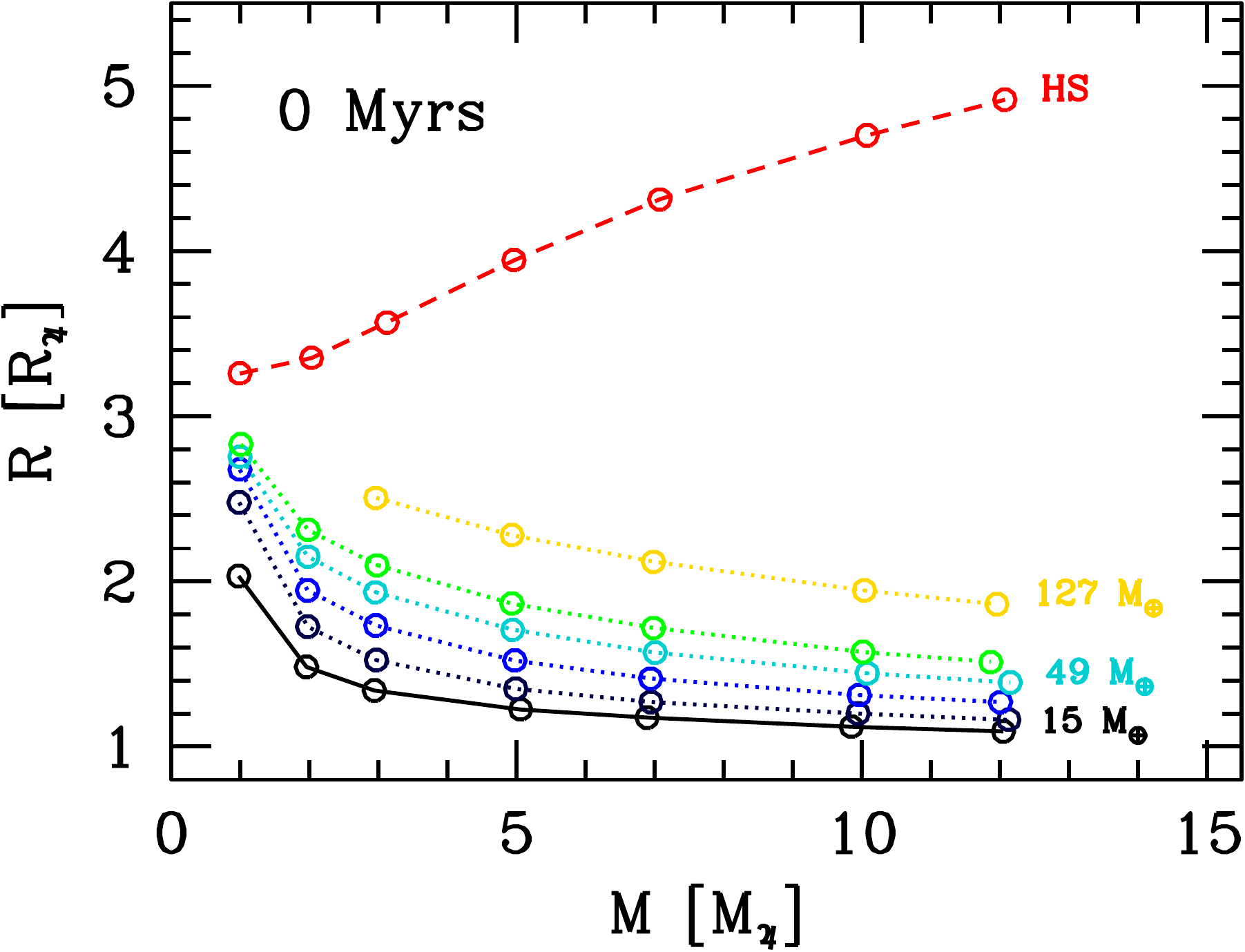}
              \includegraphics[width=1.0\textwidth]{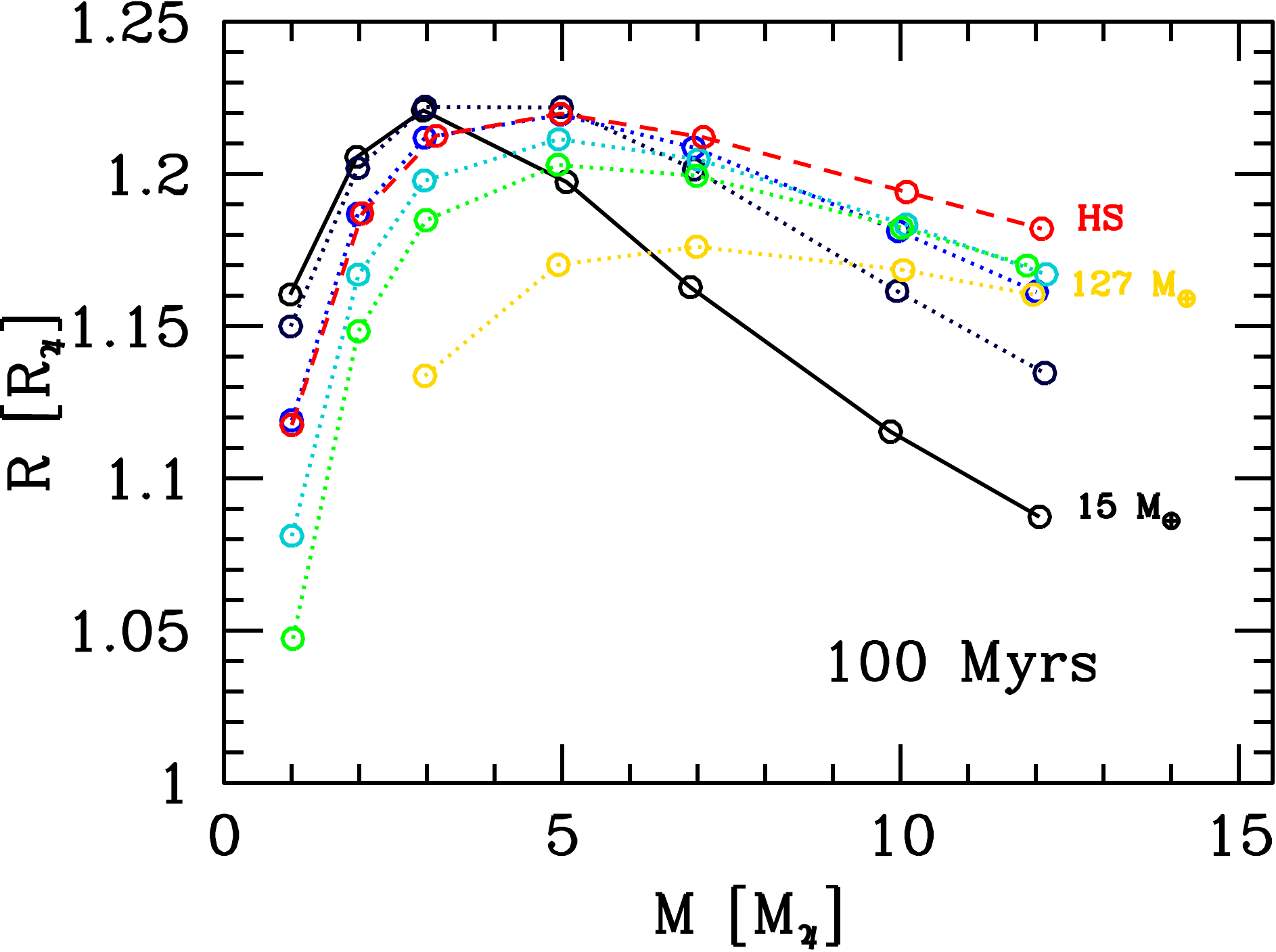}
     \end{minipage}\hfill
     \begin{minipage}{0.33\textwidth}
      \centering
       \includegraphics[width=1.0\textwidth]{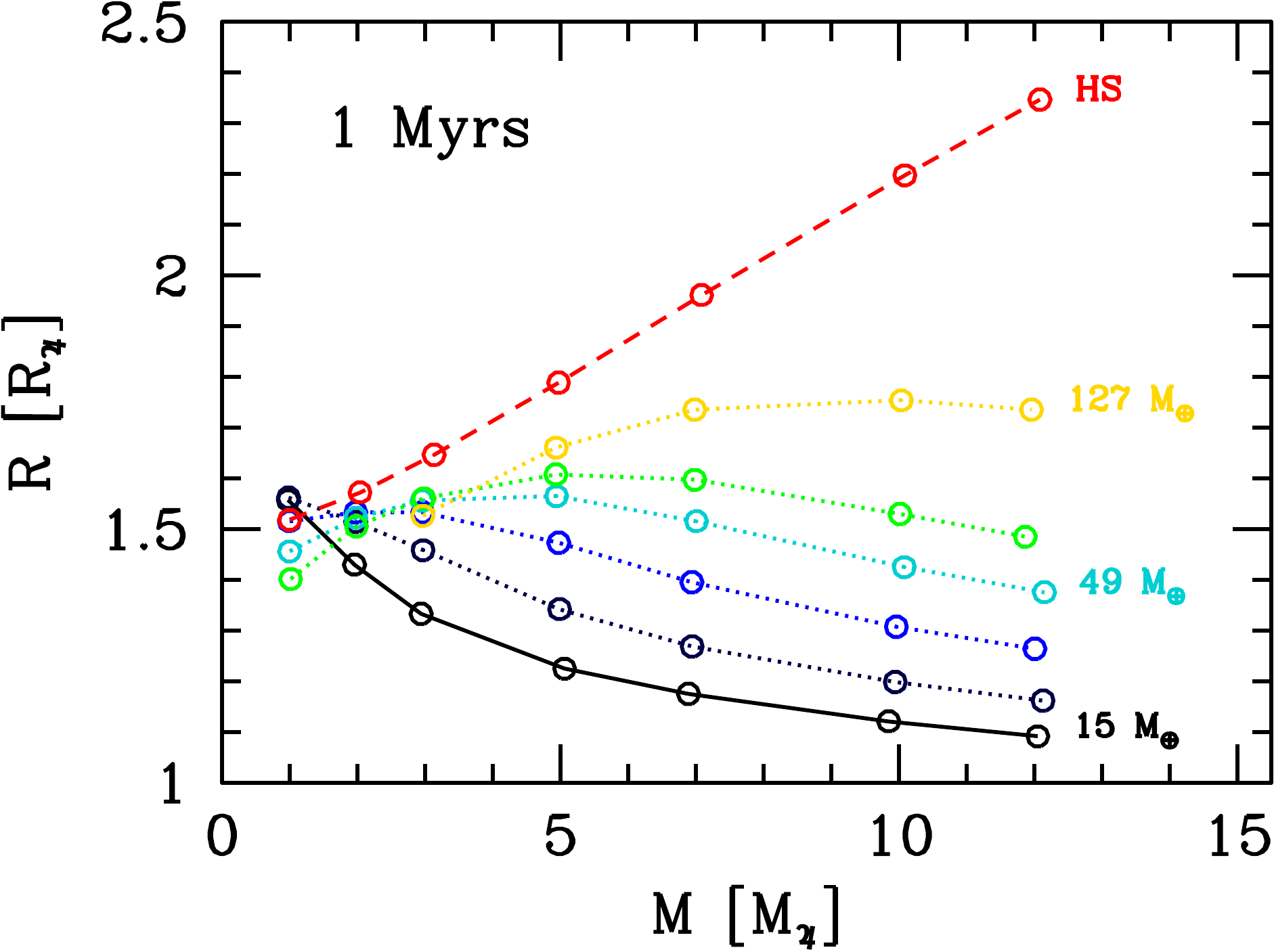}
              \includegraphics[width=1.0\textwidth]{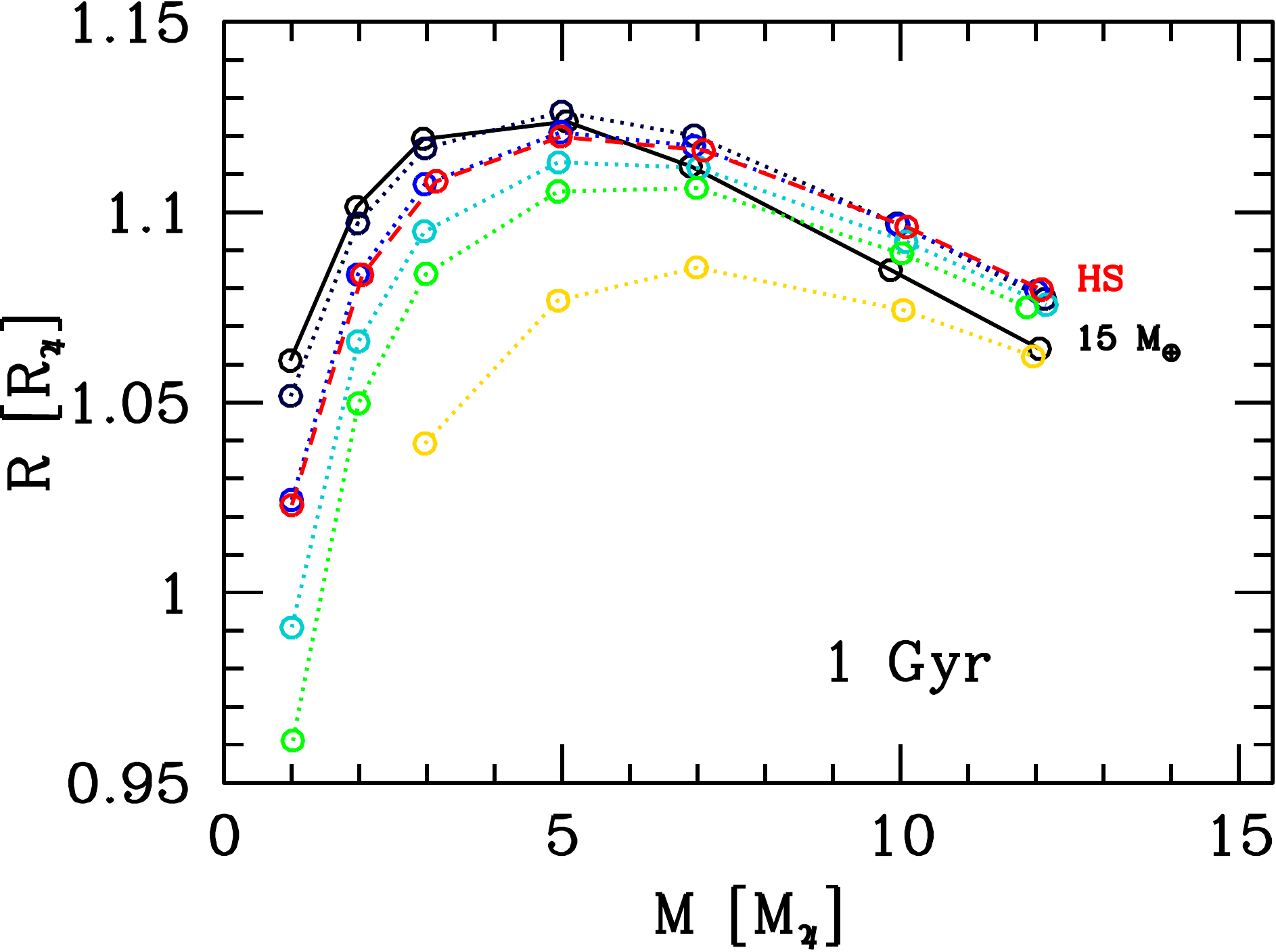}
     \end{minipage}\hfill
          \begin{minipage}{0.33\textwidth}
      \centering
       \includegraphics[width=1.0\textwidth]{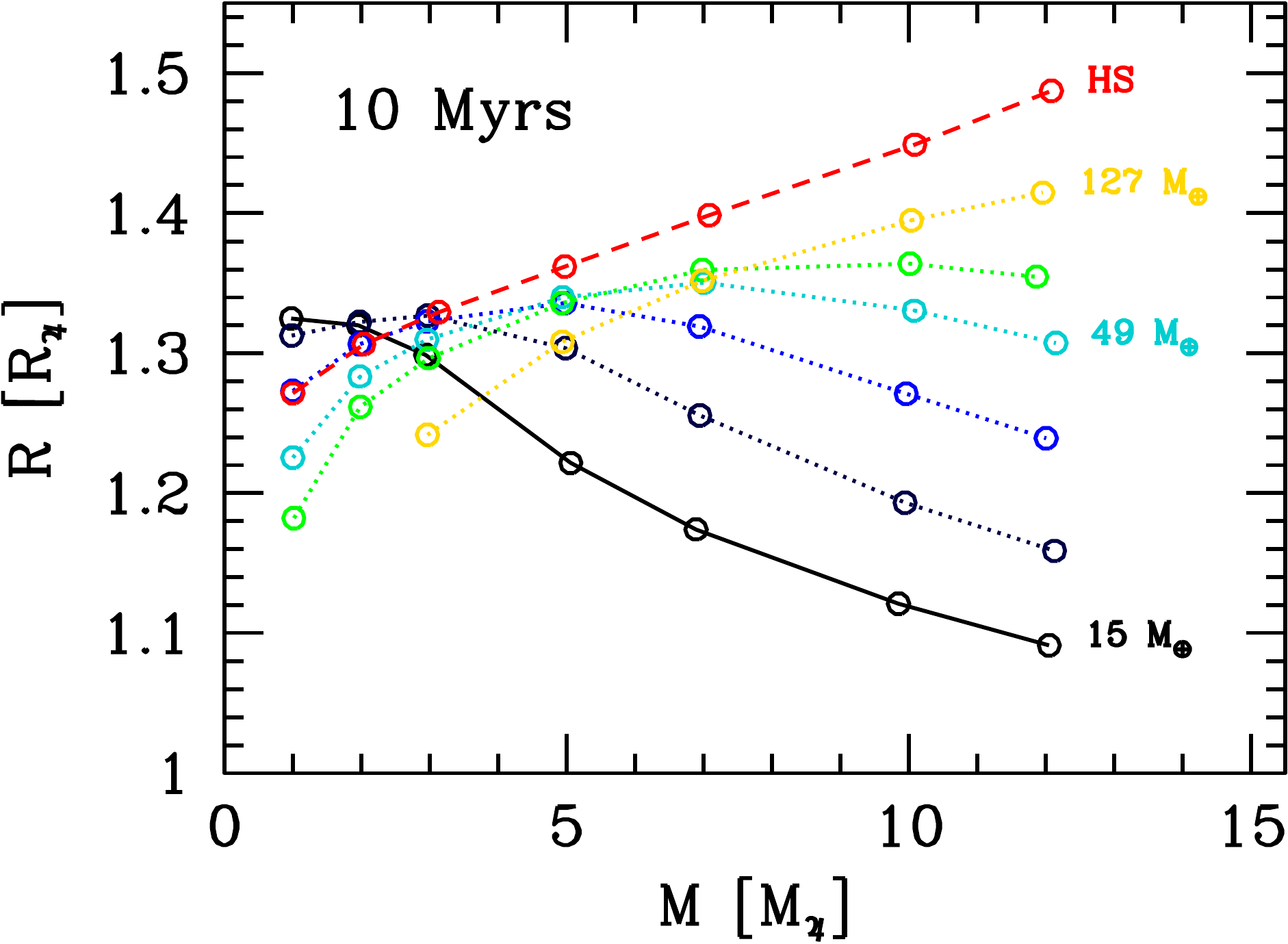}
        \includegraphics[width=1.0\textwidth]{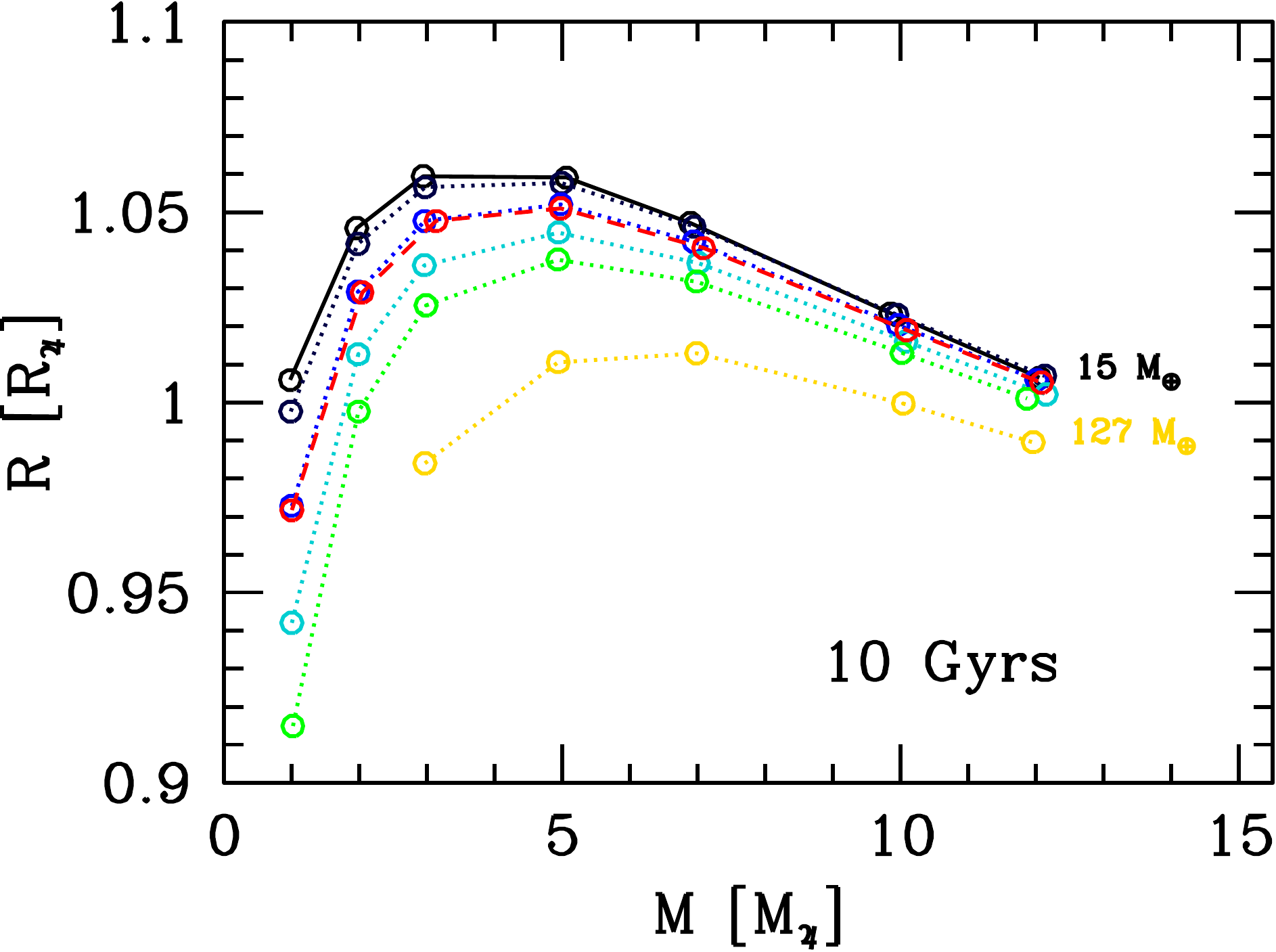}

     \end{minipage}
        \caption{Radii of planets as a function of mass between 1 and 12 $\mj$ for different core masses at six different moments in time (relative to the end of accretion) as indicated in the panels. The core masses that are also partially indicated in the panels and the associated colors are the same as in Fig. \ref{fig:fork}. HS stands again for hot accretion.  }\label{fig:radii} 
\end{figure*}

Figure \ref{fig:radii} shows the radii of the planets as a function of total mass for different core masses at six moments in time relative to the end of the formation phase. Immediately after the end of formation (which is for a fixed gas accretion rate at a later  absolute moment in time for more massive planets), there is a relatively simple situation. The planets formed with hot accretion (core mass of $\approx$35$\mearth$) have large radii that increase with increasing mass, well separated from the cold accretion planets. The initial radius of a 12 $\mj$ planet is quite large ($\approx$5 $\rj$), and also a 1$\mj$ planet is at about 3.2 $\rj$. The planets formed with cold accretion in contrast all have radii that decrease with increasing mass, a consequence of an increasingly large fraction of the planet's total mass that has passed through the entropy-reducing shock. The higher the core mass, the larger the radius, a consequence of the higher entropy for more massive cores, as we had seen in the exemplary simulations (Sect. \ref{sect:simsradius}). The higher entropy thus dominates at this time  over the well-known mechanism that a higher heavy element fraction normally leads to a smaller total radius \cite[e.g.,][]{fortneymarley2007a,baraffechabrier2008a}. 

The further evolution is governed by the interplay of the different entropy states, the different core masses, and the different Kelvin-Helmholtz timescales that are very short for hot  (high $\mcore$) low-mass planets, but very long for cold (low $\mcore$) massive planets. The KH timescale sets the time after which the planets have forgotten the initial conditions \citepalias{marleyfortney2007}. At  1 Myrs, the curves for the 1 $\mj$ planet have already crossed over, so that the planet with the most massive core now has the smallest radius, clearly a consequence of the very short $\tKH$ of this planet ($\sim$10$^{5}$ yrs). For increasingly massive planets, it takes increasingly long until  cross over occurs and the ``normal'' sequence of radii as a function of core mass is established. For the 12$\mj$ planet, e.g., cross over only happens at about 1 Gyr. At 10 Myrs, the ``normal'' order has been established also for 2$\mj$ planets, while the the radii for all 3$\mj$ planets are nearly equal with the exception of the planet with the most massive core. The radii of the hot accretion planets have significantly decreased by now, so that there is nearly a continuum of radii going at 12$\mj$ from 1.1 to 1.5$\rj$ that includes the cold and hot accretion planets.  At 100 Myrs, another feature becomes visible, namely the well-known local maximum in the planetary mass-radius relationship \citep{zapolskysalpeter1969} at about 4 $\mj$ that is due to the increasing degeneracy in the interior \citep[see, e.g.,][for a review]{chabrierbaraffe2009}. This feature becomes even better visible at 1 Gyr. At 12$\mj$, the radius of the 127 $\mearth$ core planet is now approximately equal to the one of the $15\mearth$ core planet. To reach this state, the former planet has contracted by about 0.9 $\rj$  since formation,  while the latter has only shrunk by a minuscule 0.02 $\rj$.   At 10 Gyrs, the direct influence of the core to reduce the radius eventually dominates at all masses over  its influence during formation to cause larger radii via a  higher entropy, so  that the  ``normal'' sequence of radii is established everywhere. The impact of the core mass is larger for planets with a smaller total mass, as expected \cite[e.g.,][]{fortneymarley2007a}.

\section{Conclusions}\label{sect:conclusions}
The post-formation luminosity of giant planets forming via core accretion with a radiatively efficient shock depends significantly on the mass of the core, roughly like $\mcore^{2-3}$. This dependency means that there is no single well-defined luminosity associated with core accretion even if all shock luminosity is radiated, but that there is a continuum of post-formation states (entropies, radii, and luminosities). The states range from cold states for core masses of $\sim$15$\mearth$ as in \citetalias{marleyfortney2007}, over intermediate warm  states \citepalias{spiegelburrows2012} to states that approach the classical hot state at core masses exceeding 100 $\mearth$. The timescale until the luminosity of hot and cold accretion planets agree is strongly reduced for massive cores.

Internal structure models of transiting  giant exoplanets \citep[e.g.,][]{guillotsantos2006,burrowshubeny2007,baraffechabrier2008a,millerfortney2011} indicate large amounts of heavy elements ($\sim$100$\mearth$) to be present in several planets, and some planets \citep[e.g., CoRoT-20b,][]{deleuilbonomo2012}, even seem to contain 300-1000 $\mearth$. It is unknown where these solids reside inside the planets (concentrated in the core or mixed through the envelope) and how/when they were accreted. For the influence of planetesimal accretion on the post-formation luminosity to be strong, it is necessary that (1) the solids are accreted in large amounts during or before runaway gas accretion, and (2) that they sink deep into the planet during this period. In this paper, we do not make any  claims whether or not this is possible for an, e.g.,  100$\mearth$ core in particular if the planet forms at a large orbital distance ($\gg$10 AU) where the accretion timescale at least for km-sized planetesimals is slow. We note that for smaller objects this seems to be significantly different, since aerodynamically assisted accretion  rates of boulders or pebbles can be very high also in the outer disk \citep[e.g.,][]{ormelklahr2010,lambrechtsjohansen2012}. In any case, future studies must investigate both requirement based on solid accretion rates and the fate of solids inside the planet \citep[cf.][]{iaroslavitzpodolak2007}. We just point out that if massive cores can indeed form, then their impact on the luminosity is large, and this already for a more ``normal'' core mass of 30 $\mearth$. 

A consequence of our results is that it becomes difficult to rule out core accretion as formation mechanism based solely on luminosity  for directly imaged planets that are more luminous than predicted by \citetalias{marleyfortney2007} (who considered $\mcore\leq19\mearth$) or the equivalent simulations here for low core masses. Instead of invoking gravitational instability as the consequently necessary formation mechanism, the high luminosity could also be caused, at least in principle, by more massive cores in some of the planets.  Such a more massive core has similar consequences as an accretion shock that does not radiate all accretion shock luminosity \citepalias{spiegelburrows2012}, meaning that there is a degeneracy between the two mechanism (only at later times, the two become distinguishable through a smaller total radius of a planet with a more massive core). 

In summary we see that concerning the luminosity of young Jupiters, we  face a situation that is more complex than previously thought. The formation mechanism, the shock structure, but also the core mass play an important role. The atmospheric composition measured with multi-band photometry or spectroscopy \citep[e.g.,][]{jansonbergfors2010,bonnefoyboccaletti2013,konopackybarman2013} could be helpful in this situation to distinguish the different scenarios, because it seems likely that the presence of a massive core is associated with an enhanced metal content of the atmosphere.

\acknowledgements{I thank Hubert Klahr, Paul Molli\`ere, Yann Alibert, Willy Benz,  Micka\"el Bonnefoy, Thomas Henning and Kai-Martin Dittkrist  for useful discussions, {and the Max-Planck-Gesellschaft for the Reimar-L\"ust Fellowship}. Computations were made on the BATCHELOR cluster at MPIA. {I thank the referee Dr. David Spiegel for a constructive review. During the preparation of this
work, we became aware that \citet{bodenheimerdangelo2013} independently also found
 the dependency of the post-formation entropy on the core mass in
another context (deuterium burning). For a 12 $\mj$ planet and in the overlapping core mass range of 15 to 35 $\mearth$, the two works predict post-formation entropies as  function of the core mass that agree well, namely to about 0.2 $k_{\rm B}$/baryon (Fig. 11 in  \citet{bodenheimerdangelo2013} and Fig. \ref{fig:fork} of this work).  } }

\appendix

\section{Power-law approximation for the post-formation luminosity}\label{sect:l0mcore}
\begin{figure}[tb]
\begin{center}
\includegraphics[width=0.95\columnwidth]{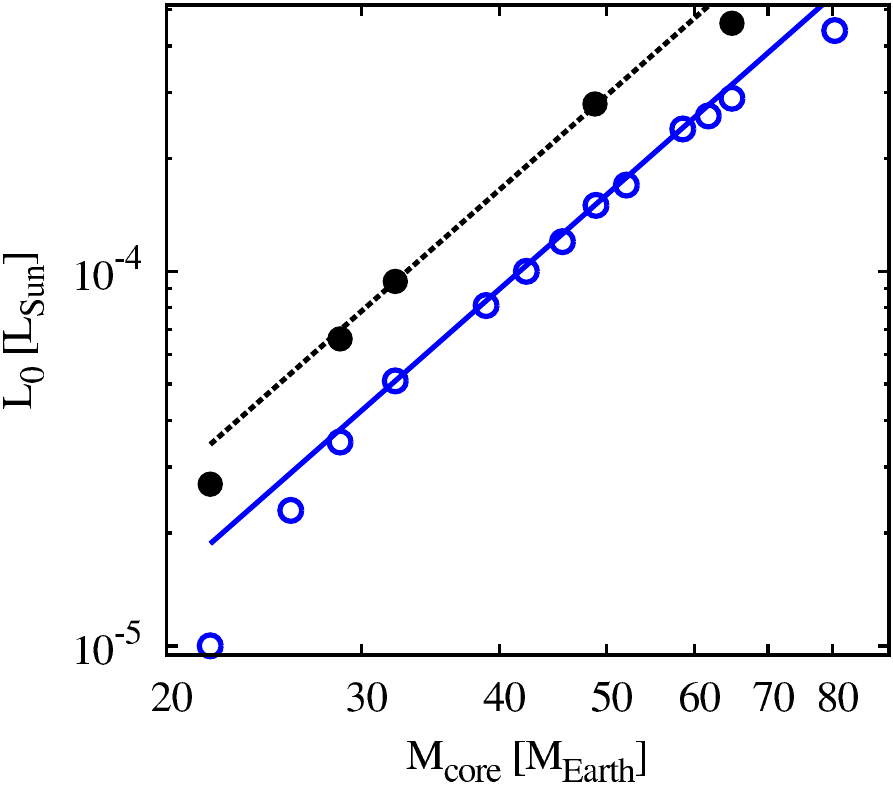}
\caption{{Post-formation luminosity of planets with a total mass of 5 and 10 $\mj$ (black filled and blue open points, respectively), as a function of the mass of the core. The  black dashed and blue solid lines show the corresponding power-law approximation for intermediate core masses, scaling as $\mcore^{2.6}$.}}
\label{fig:l0mcore510}
\end{center}
\end{figure}

{In Sect. \ref{sect:lumi4sims} an approximative power-law dependency of the post-formation luminosity on the core mass was mentioned for 10 $\mj$ planets and core masses between 32 and 59 $\mearth$. Figure \ref{fig:l0mcore510} shows the post-formation luminosity $L_{0}$ for planets with total masses of 5 and 10 $\mj$ and core masses between 22 and 80 $\mearth$. The lines show the corresponding power-law approximation with a scaling as  $\mcore^{2.6}$. The power-laws were normalized at the luminosity of the planet with a core mass of approximately 33 $\mearth$. }

{We first note again the typical ``luminosity inversion'', i.e., the fact that for cold accretion, the luminosity directly after the end of the formation phase is lower for a more massive planet (\citetalias{marleyfortney2007}), as also (partially) visible in Fig. \ref{fig:mammamia}. The plot further shows that the power-law approximation agrees well with the numerical results at intermediate core masses (roughly 30 to 50 $\mearth$). For larger core masses, the increase of $L_{0}$ with increasing $\mcore$ is in contrast weaker than predicted by the fit. For core masses less than 30 $\mearth$, the numerical results also diverge  from the power-law. At these low core masses, the dependency on $\mcore$ is clearly stronger than $\mcore^{2.6}$, so that the fit would predict too high $L_{0}$. This is (at least quantitatively) in agreement with the results of  \citetalias{marleyfortney2007}: For a 10 $\mj$ planet with a core mass of 19 $\mearth$, they find a $L_{0}$ of $2-3\times10^{-6}\lsun$, while the power-law would predict a luminosity of $1.3\times10^{-5}\lsun$.}
 
\section{Impact of model settings on the core mass - initial entropy mapping}\label{sect:modelsettingsMcoreS0}
{As mentioned in Section \ref{sect:model}, the outer boundary conditions  in this work differ from the ones in \citet{bodenheimerhubickyj2000} by the additional photospheric pressure. In this section, we investigate the impact of this model setting, and additionally of the opacity due to grains in the protoplanetary envelope. The goal is  to gain insight in the robustness of the core mass - initial entropy mapping within our our 1D model framework (cf. Sect. \ref{sect:entropy}).  }

\begin{figure}
\begin{center}
\includegraphics[width=0.95\columnwidth]{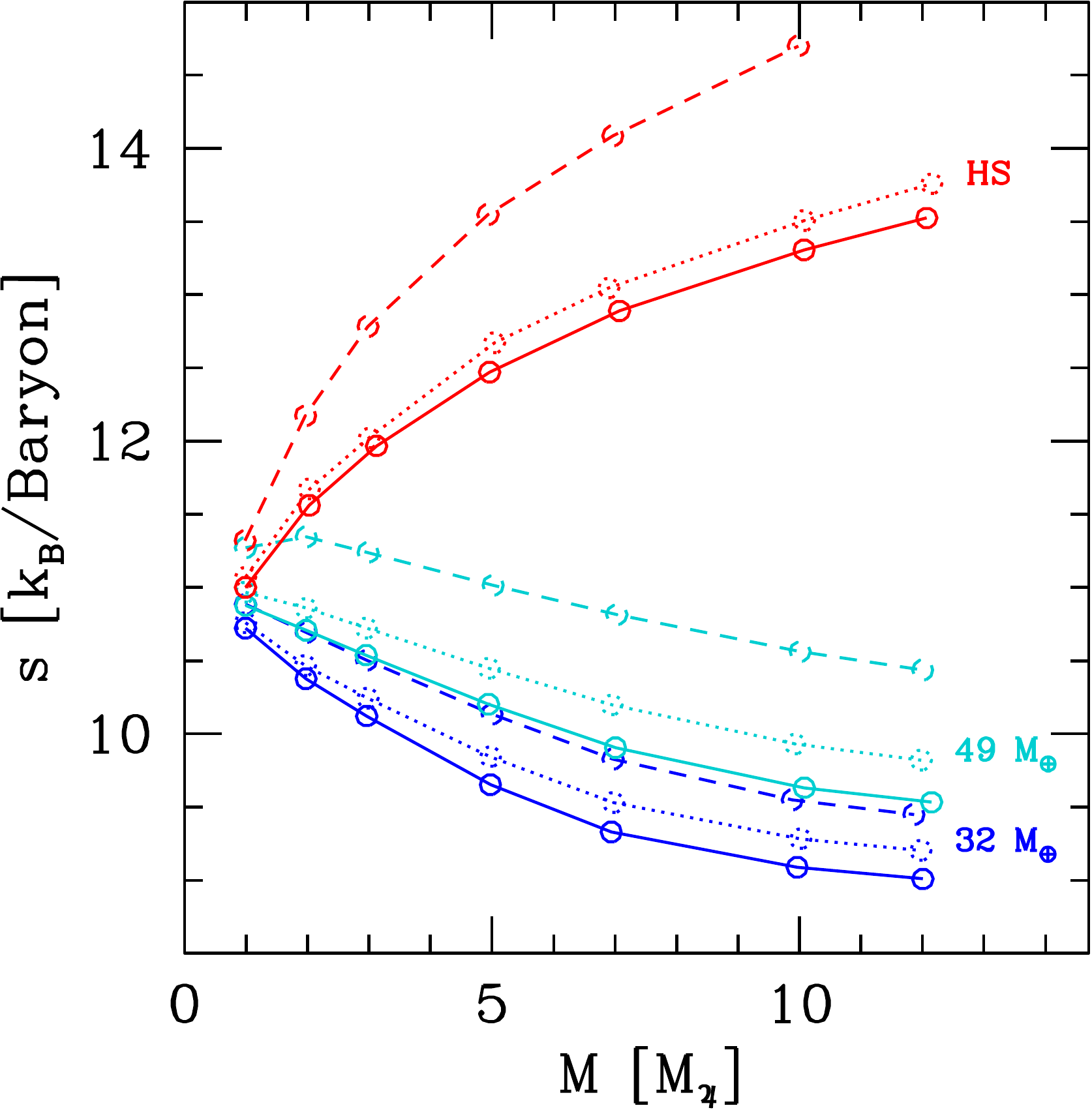}
\caption{{Post-formation entropy as function of total mass. The dark (light) blue lines are for cold accretion and a core mass of 32 (49) $\mearth$. The red curves are for hot accretion. The solid lines are the nominal model and identical to Fig. \ref{fig:fork}. The dotted curves are simulations without the photospheric pressure. The dashed lines are for a full ISM grain opacity.}}
\label{fig:forkpeddfopa}
\end{center}
\end{figure}

{Figure \ref{fig:forkpeddfopa} shows the post-formation entropy as a function of  mass. The solid lines are the same as in Fig. \ref{fig:fork} and show the entropy for cold accretion and core masses of 32 and 49 $\mearth$, and additionally for hot accretion. The dotted lines show the entropies for corresponding simulations without the photospheric pressure in the outer boundary conditions but otherwise identical settings. We see that the omission of this term leads to initial entropies that are for massive planets about 0.2 to 0.3 higher, which is a rather small change. For lower-mass planets, the impact is even smaller.  A generally higher entropy is expected, because with the ram pressure only, the pressure in the outermost layers is smaller, while the temperature is nearly identical. The plot also makes clear that the general shape of the curves remains quite similar (for cold accretion, decrease of $\s0$ with $M$ and increase with $\mcore$), so that the qualitative result is unchanged under  a modification of these model settings.  }

{We have also tested the impact of the opacity due to grains suspended in the protoplanetary envelope. Inspired by the work of \citet{movshovitzbodenheimer2010}, we have reduced  the grain opacity by a factor $\fopa$=0.003 relative to the ISM in the nominal simulation \citep[cf.][]{mordasinialibert2012a}. For comparison, \citetalias[][]{marleyfortney2007} used a reduction factor of 0.02. \citet{movshovitzbodenheimer2010} studied the evolution of the grains during  phase II of giant planet formation, therefore it is  not clear if their results of  can be applied to the runaway gas accretion phase of high mass planets we study here. To quantify the impact of this assumption, the dashed lines show the post-formation entropy for simulations with full ISM opacity and otherwise nominal settings. The qualitative result is the same as for the photospheric pressure: at full opacity, higher entropies result. The quantitative impact of a full ISM opacity is however larger, with difference of up to about 1.0 in $\s0$ (for cold accretion), which is significant. This is a consequence of the slower contraction of the radius at a higher opacity, which makes that the planet has at any given mass during the detached phase a larger radius, meaning that less gravitational potential energy is radiated at the accretion shock.  Depending on the temperature, a higher grain opacity also leads to a smaller photospheric pressure, which increases the entropy, as seen before. We have finally also found a certain inverse correlation of the distance from where the gas is assumed to fall onto the planet (infinity, $\rhill$, $\rhill/3$) and $\s0$.} 

{Qualitatively, the  general correlations of $M$, $\mcore$ and $S_0$ found in this work  remain under these various changes of the model assumptions, meaning that the general trend appears to be relatively robust within our model framework. The  quantitative values of the post-formation entropy are in contrast dependent in a non-negligible way on specific assumptions even within the 1D framework. If the general correlations presist also in 3D remains to be shown in future work.} 
 
\bibliographystyle{aa} % style aa.bst 
\bibliography{biball2013new.bib}

\end{document}